\documentclass[12pt]{article}
\usepackage[utf8]{inputenc}
\usepackage[english]{babel}
\usepackage[centertags]{amsmath}
\usepackage{amssymb}
\usepackage{bm}
\usepackage{amsbsy}
\usepackage{graphicx}
\usepackage{appendix}
\usepackage{mathrsfs}
\usepackage{epsfig}
\usepackage{color}
\usepackage{euscript}
\usepackage{ulem}
\usepackage{multirow}
\usepackage{cite}
\usepackage{textcomp}

\definecolor{darkraspberry}{rgb}{0.53,0.15,0.34}

\definecolor{darkblue}{rgb}{0,0,1}

\definecolor{dgreen}{rgb}{0,0.6,0}

\definecolor{darkred}{rgb}{0.8,0,1}

\usepackage{jheppubm}

\newcommand{\nn}{\nonumber}
\newcommand{\be}{\begin{equation}}
\newcommand{\ee}{\end{equation}}
\newcommand{\bea}{\begin{eqnarray}}
\newcommand{\eea}{\end{eqnarray}}

\newcommand{\fb}{\mathfrak{b}}

\newcommand{\cA}{\cal A}
\newcommand{\cL}{\cal L}

\numberwithin{equation}{section}

\title{Magnetic Catalysis in Holographic Model with Two Types of
  Anisotropy for Heavy Quarks} 

\author{Irina Ya. Aref'eva$^a$, Ali Hajilou$^a$, Kristina Rannu$^b$
  and Pavel Slepov$^a$} 

\affiliation{$^a$Steklov Mathematical Institute, Russian Academy of
  Sciences, \\ Gubkina str. 8, 119991, Moscow, Russia  \\ 
  $^b$Peoples Friendship University of Russia, Miklukho-Maklaya
  str. 6, 117198, Moscow, Russia} 

\emailAdd{arefeva@mi-ras.ru}
\emailAdd{hajilou@mi-ras.ru}
\emailAdd{rannu-ka@rudn.ru}
\emailAdd{slepov@mi-ras.ru}

\abstract{In our previous paper \cite{ARS-Heavy-2020} we have
  constructed a twice anisotropic five-dimensional holographic model
  supported by Einstein-dilaton-three-Maxwell action that reproduced
  some essential features of the ``heavy quarks'' model. However, that
  model did not describe the magnetic catalysis (MC) phenomena
  expected from lattice results for the QGP made up from heavy
  quarks. In this paper we fill this gap and construct the model that
  improves the previous one. It keeps typical properties of the heavy
  quarks phase diagram, and meanwhile  possesses the MC. The
  deformation of previous model includes the modification of the
  ``heavy quarks'' warp factor and the coupling function for the
  Maxwell field providing the non-trivial chemical potential.}


\begin{document}

\maketitle


\newpage

\section{Introduction}

Quantum chromodynamics (QCD) is a theory that describes strong
interactions between subatomic particles such as quarks and
gluons. Complete description of the QCD phase diagram in a parameter
space with temperature, chemical potential, quark masses, anisotropy,
magnetic field etc. is a challenging and very important task in high
energy physics. Standard methods to do calculations in QCD such as
perturbation no longer work for the strongly coupled regime of this
theory, while the lattice theory has problems with non-zero chemical
potential calculations. Hence, to understand physics of the strongly
coupled quark–gluon plasma (QGP) produced in heavy ion collisions
(HIC) at RHIC and at the LHC, and future experiments, we need a
non-perturbative approach \cite{Casalderrey-Solana:2011dxg,
  Arefeva:2014kyw, Arefeva:2021kku}.

According to the results of the experiments with relativistic HIC, it
is believed  that a very strong magnetic field, $eB \sim 0.3$ GeV$^2$,
is created in the early stages of the collision \cite{Skokov:2009qp,
  Voronyuk:2011jd, Bzdak:2011yy, Deng:2012pc}. Therefore, magnetic
field is an important parameter characterizing the QCD phase diagram 
expected from the experiments with relativistic HIC
\cite{Miransky:2015ava, Kharzeev:2012ph, Andersen:2014xxa}. Studying
QCD in the background of magnetic field has received much attention
recent years, among other things, because of such an interesting
phenomena as chiral magnetic effect \cite{Fukushima:2008xe,
  Kharzeev:2007jp}, magnetic catalysis (MC) \cite{Shovkovy:2012zn,
  Miransky:2002rp}, inverse magnetic catalysis (IMC)
\cite{Mao:2016fha, Fukushima:2012kc}, as well as the early Universe  
physics \cite{Grasso:2000wj, Vachaspati:1991nm} and dense neutron
stars \cite{Duncan:1992hi}. Therefore, investigation of the magnetic
field effect on the QCD features and in particular its phase diagram
is very interesting and crucial for better understanding of the
QCD. Besides, such an investigation has been considered via lattice
calculations \cite{DElia:2010abb, DElia:2018xwo, Bali:2012zg,
  Bali:2013esa}. For more information and detailed reviews the 
interested reader is referred to \cite{Miransky:2002rp,
  Kharzeev:2012ph} and references therein. In particular, many
holographic QCD models have been developed to investigate the effect
of magnetic field on the characteristics of QCD \cite{Johnson:2008vna,
  Mamo:2015dea, Rougemont:2015oea, Dudal:2015wfn, Li:2016gfn,
  Dudal:2016joz, Rodrigues:2017cha, Rodrigues:2017iqi,
  Rodrigues:2018pep, Gursoy:2017wzz, Bohra:2019ebj, Dudal:2021jav,
  Arefeva:2022bhx, Arefeva:2021mag, Rannu:2022fxw, Arefeva:2023ter,
  Jena:2022nzw, Shukla:2023pbp, Colangelo:2021kmn}.

The enhancement effect of the phase transition temperature under
the magnetic field increasing is known as MC phenomenon, the opposite
effect is called IMC. Lattice calculations show that for small
chemical potential there is a substantial influence of the magnetic
field on the QCD phase diagram structure. This influence essentially
depends on the quark mass: for small quark mass (light quarks) IMC
takes place, meanwhile for large mass (heavy quarks) MC occurs. In
this context note that lattice calculations predict different types of
phase transitions even for small chemical potential and zero magnetic
field -- we have a crossover for light quarks, and a first-order phase
transition for heavy quarks. The holographic QCD models for heavy and
light quarks constructed  in \cite{Brown:1990ev, AR-2018,
  ARS-Light-2020, Li:2017tdz, Yang:2014bqa, Arefeva:2022bhx} reproduce
these phase diagram features at small chemical potential and predict
new interesting phenomena for finite chemical potential, in
particular, the locations of the critical end points. In our previous
papers \cite{ARS-Light-2022}, see also \cite{Li:2016gfn}, where the
light quark  holographic model with non-zero magnetic field is 
investigated, it has been shown that IMC takes place. Our paper
\cite{ARS-Heavy-2020} shows that the heavy quark holographic model 
\cite{AR-2018} still has IMC, not MC, that contradicts with lattice
zero chemical potential calculation. This indicates that one has to
modify the heavy quark holographic model~\cite{ARS-Heavy-2020}. 
 
In the current paper we fill this gap and construct a heavy quark
model that improves the previous one \cite{AR-2018, ARS-2019,
  ARS-Heavy-2020}. The main goal of the improvements is to get the MC
phenomenon in holographic description of the heavy quarks' first order
phase transition scenario with external magnetic field keeping typical
properties of the heavy quarks phase diagram. For this purpose we can
consider additional $z^4$- \cite{White:2007tu, Pirner:2009gr,
  He:2020fdi} or/and $z^5$-terms \cite{Bohra:2020qom, Rannu-2024} into
the exponent warp factor. In particular, within this holographic model
we show that $z^4$-term allows to produce the MC phenomenon
required. 

As we have emphasized in the previous papers \cite{Arefeva:2014vjl,
  AR-2018,ARS-Light-2020}, there is a reason to introduce one more
parameter characterizing the QCD phase diagram -- an anisotropy
parameter $\nu$. Non-central HIC produces anisotropic QGP, and the
isotropisation time is estimated as $1$--$5$ fm/$c \sim 10^{-24}$ s
\cite{Strickland:2013uga}. Anisotropic holographic models have been
used to study QGP in \cite{Mateos:2011ix, Mateos:2011tv, Janik:2008tc,
  Rebhan:2012bw, Giataganas:2012zy, Arefeva:2014vjl, Ageev:2016gtl,
  AGG, Arefeva:2016rob, Giataganas:2017koz, Gursoy:2018ydr}. One of
the main purposes to consider anisotropic models is to describe the
experimental energy dependence of total multiplicity of particles
created in HIC \cite{ALICE:2015juo}. In \cite{Arefeva:2014vjl} it has
been shown that the choice of the primary anisotropy parameter value
about $\nu = 4.5$ reproduces the energy dependence of total
multiplicity \cite{ALICE:2015juo}. Note that isotropic models could
not reproduce it (for more details see \cite{Arefeva:2014vjl} and
references therein). In addition, it is very interesting to know how
the primary (spatial) anisotropy can affect the QCD phase transition
temperature. Note also that there is another type of anisotropy due
to magnetic field and its effect on the QCD phase diagram is a subject
of interest. 

In this work we set up a twice anisotropic ``heavy quarks'' model. In
fact, we consider 5-dim Einstein-Maxwell-dilaton action with three
Maxwell fields: the first Maxwell field sets up finite non-zero
chemical potential in the gauge theory, the second Maxwell field
provides the primary spatial anisotropy to reproduce the multiplicity
dependence on energy, and the 3-rd Maxwell field provides another
anisotropy that originates from magnetic field in the gauge theory. We
use an anisotropic metric as an ansatz to solve Einstein equations and
the field equations self-consistently. The central question of the
current investigation is the form of the warp factor able to provide
the MC phenomenon within the constructed holographic model. This our
consideration shows a phenomenological character of the bottom-up
holographic models \cite{Erlich:2005qh, Andreev:2006nw, Gursoy:2010fj,
  Colangelo:2011sr, Arefeva:2019yzy, Li:2013oda, Li:2012ay,
  Gursoy:2008za, Arefeva:2020uec, Arefeva:2021kku, Mia:2010zu,
  Dudal:2018ztm, Dudal:2017max, Arefeva:2020vhf, Yang:2014bqa,
  Li:2014dsa, Yang:2015aia, Chelabi:2015cwn, Fang:2015ytf,
  Arefeva:2018jyu, Fang:2018axm, Chen:2018msc, Chen:2019rez, AR-2018,
  ARS-2019, ARS-Heavy-2020, ARS-Light-2020, Arefeva:2021jpa,
  ARS-Light-2022, Hajilou:2021wmz, He:2020fdi, Bohra:2020qom,
  He:2013qq, He:2010ye, Gursoy:2007er, Gursoy:2009jd, Gursoy:2007cb,
  Gursoy:2008bu, Karch:2006pv}, that is different from the top-down
holographic models \cite{Sakai:2004cn, Sakai:2005yt, Karch:2002sh,
  Kruczenski:2003be, Kruczenski:2003uq}. 

This paper is organized as follows. In section \ref{model5d} we
present a 5-dim holographic model to describe a hot dense anisotropic 
QCD in the magnetic field background. In section \ref{mcmodel} we
introduce an appropriate  warp factor able to produce MC phenomenon in
this holographic model and obtain the first order phase transition for
the model parameters. In section \ref{results} we review our main
results. This work in complemented with Appendix \ref{eeom} where we 
solve EOMs, Appendix \ref{blackf} where we present expressions for
the blackening function derivatives, gauge coupling functions and
dilaton potential, and Appendix \ref{appC} where we consider the
relation of our setting with the setting \cite{He:2020fdi}
explicitly.

\section{Holographic Model with three Maxwell Fields} \label{model5d}

Let us take the Lagrangian in Einstein frame used in
\cite{ARS-Heavy-2020}:
\begin{gather}
  {\cL} = \sqrt{-g} \left[ R 
    - \cfrac{f_0(\phi)}{4} \, F_0^2 
    - \cfrac{f_1(\phi)}{4} \, F_1^2
    - \cfrac{f_3(\phi)}{4} \, F_3^2
    - \cfrac{1}{2} \, \partial_{\mu} \phi \, \partial^{\mu} \phi
    - V(\phi) \right], \label{eq:2.01}
\end{gather}
where $R$ is Ricci scalar, $\phi$ is the scalar field, $f_0(\phi)$,
$f_1(\phi)$ and $f_3(\phi)$ are the coupling functions associated with
stresses $F_0$, $F_1$ and $F_3$ of Maxwell fields, and $V(\phi)$ is
the scalar field potential. In this paper we considered $F_0$, $F_1$
and $F_3$ as first, second and third Maxwell fields, respectively.

Varying Lagrangian (\ref{eq:2.01}) over the metric we get Einstein
equations of motion (EOMs):
\begin{gather}
  G_{\mu \nu} =T_{\mu \nu}, \label{EEOM}
\end{gather}
where
\begin{gather}
  G_{\mu \nu} = R_{\mu \nu} - \cfrac{1}{2} \, g_{\mu \nu} R, \qquad
  \cfrac{\delta S_m}{\delta g^{\mu \nu}} = \frac{1}{2} \, T_{\mu \nu}
  \sqrt{-g}, \label{eq:2.06}
\end{gather}
and varying over the fields gives the fields equations
\begin{gather}
 -\nabla_{\mu}\nabla^{\mu} \phi  + V'(\phi) + \!\! \sum_{i=0,1,3} \!\!
  \cfrac{f_i'(\phi)}{4} \, F_{(i)}^2 = 0, \label{phiEOM} \\
  \partial_{\mu} \left( \sqrt{-g} \, f_i \, F_{(i)}^{\mu \nu} \right)
  = 0. \label{EMEOM}
\end{gather}

Let us take the metric ansatz in the following form:
\begin{gather}
  ds^2 = \cfrac{L^2}{z^2} \ \fb(z) \left[
    - \, g(z) \, dt^2 + dx_1^2 
    + \left( \cfrac{z}{L} \right)^{2-\frac{2}{\nu}} dx_2^2
    + e^{c_B z^2} \left( \cfrac{z}{L} \right)^{2-\frac{2}{\nu}} dx_3^2
    + \cfrac{dz^2}{g(z)} \right] \! , \label{eq:2.04} \\
  \fb(z) = e^{2{\cA}(z)}, \nn
\end{gather}
and for matter fields\footnote{Also, we can add a new Maxwell field
  $F_2$ with magnetic ansatz $F_2 = q_2 \, dx^1 \wedge dx^3$ to our
  model.}
\begin{gather}
  \phi = \phi(z), \quad \, \label{eq:2.02} \\
  \begin{split}
    F_0\, -\, \mbox{electric ansatz } \,\, A_0 &= A_t(z), \quad 
    A_{i}=0,\,\,i= 1,2,3,4, \\
    F_k\,- \, \mbox{magnetic ansatz} \quad
    F_1 &= q_1 \, dx^2 \wedge dx^3, \quad 
    F_3 = q_3 \, dx^1 \wedge dx^2\, . 
  \end{split}\label{eq:2.03}
\end{gather}
In \eqref{eq:2.04} $L$ is the AdS-radius, $\fb(z)$ is the warp factor
set by ${\cA}(z)$, $g(z)$ is the blackening function, $\nu$ is the
parameter of primary anisotropy caused by non-symmetry of heavy-ion
collision (HIC), and $c_B$ is the coefficient of secondary anisotropy
related to the magnetic field $F_3$. Choice of ${\cA}(z)$ determines
the heavy/light quarks description of the model. In previous works we
considered ${\cA}(z) = - \, c z^2/4$ for heavy quarks \cite{AR-2018,
  ARS-2019, ARS-Heavy-2020} and ${\cA}(z) = - \, a \, \ln (b z^2 + 1)$
for light quarks \cite{ARS-Light-2020, ARS-Light-2022}. In
\eqref{eq:2.03} $q_1$ and $q_3$ are constant ``charges''.

The explicit form of the EOM (\ref{eq:2.06}--\ref{EMEOM}) with ansatz
(\ref{eq:2.02})--(\ref{eq:2.03}) is given in Appendix
(\ref{eq:2.16}--\ref{eq:2.22}). Investigation of their
self-consistency shows that there is one dependent equation in the
system and all other equations are independent. Thus, system
(\ref{eq:2.16}--\ref{eq:2.22}) is self-consistent and the dilaton
field equation (\ref{eq:2.16}) serves as a constraint.

It is important to note that the coupling function $f_0$ is defined
from the requirement to reproduce  the Regge trajectories. Functions
$f_1$ and $f_3$  are obtained from the EOM, and we find that they are
different (see Appendices \ref{eeom} and \ref{blackf} for more
details). If we take $f_0 = f_1 = f_3$, then we cannot construct a
solution of EOM within our ansatz.
 
Note that while solving  equations of motions we do not actually get
$f_0(\phi)$, $f_1(\phi)$, $f_3(\phi)$ and $V(\phi)$ dependencies, but
obtain $f_0(\phi(z)) = f_0(z)$, $f_1(\phi(z)) = f_1(z)$, $f_3(\phi(z))
= f_3(z)$ and $V(\phi(z)) = V(z)$. The reason is that
$\phi(z)$-expression is rather complicated, so the analytic expression 
for the inverse function $z = z(\phi)$ can't be written down. There
still remains the possibility to get this function via approximation,
but such a result can be useful for a limited number of aspects only
because of lack of accuracy.


\section{Magnetic Catalysis for Heavy Quarks} \label{mcmodel}

Our goal is to generalize solution \cite{ARS-Heavy-2020} to get
magnetic catalysis effect on the heavy quarks version of the phase
diagram. For this purpose we choose the deformation of the warp factor
from \cite{ARS-Heavy-2020} again. This factor has been used in
\cite{White:2007tu, Pirner:2009gr} to reproduce Cornell potential and
in \cite{He:2020fdi} also to get magnetic catalysis effect.



\subsection{Solution and thermodynamics for ${\cA}(z) = - \, cz^2/4 \,
  - p z^4$} 
 Our strategy to solve 
the EOMs presented in Appendix \ref{eeom} with the factor 
${\cA}(z) = - \, cz^2/4 \,
  - p z^4$ is the same as in \cite{ARS-Heavy-2020} and 
  \cite{ARS-Light-2020}. Subtracting (\ref{eq:2.21})
from (\ref{eq:2.20}) we get the expression for the third Maxwell
field's coupling function
\begin{gather}
  f_{3} = 2 \left( \cfrac{L}{z} \right)^{\frac{2}{\nu}} \fb g \
  \cfrac{c_B z}{q_3^2} \left(
    \cfrac{g'}{g} + \cfrac{3 \fb'}{2 \fb} - \cfrac{2}{\nu z} + c_B z
  \right) \label{eq:4.21}
\end{gather}
and rewrite equation \eqref{eq:2.18} as
\begin{gather}
  g'' + g' \left(
    \cfrac{3 \fb'}{2 \fb} - \cfrac{\nu + 2}{\nu z} - c_B z
  \right) 
  - 2 g \left(
    \cfrac{3 \fb'}{2 \fb} - \cfrac{2}{\nu z} + c_B z
  \right) c_B z
  - \left( \cfrac{z}{L} \right)^2 \cfrac{f_0 (A_t')^2}{\fb} 
  = 0. \label{eq:4.22}
\end{gather}
To derive the exact solutions we just need to specify the warp
factor:
\begin{gather}
  \fb(z) = e^{2{\cA}(z)} =  e^{ - \, c z^2/2 \, - \,2 p
    z^4}. \label{eq:4.23}
\end{gather}
Following \cite{He:2013qq, Li:2017tdz, He:2020fdi} we take $c = 4
R_{gg}/3$, $R_{gg} = 1.16$, $p = 0.273$ (this choice is dictated by
the Regge spectra and lattice QCD fitting) and solve system
(\ref{eq:2.16})--(\ref{eq:2.22}) with usual boundary conditions 
\begin{gather}
  A_t(0) = \mu, \quad A_t(z_h) = 0, \label{eq:4.24} \\
  g(0) = 1, \quad g(z_h) = 0, \label{eq:4.25} \\
  \phi(z_0) = 0, \label{eq:4.26}
\end{gather}
where $z_0$ serves to fit the string tension behavior
\cite{ARS-Light-2020}.

\begin{figure}[t!]
  \centering
  \includegraphics[scale=1]{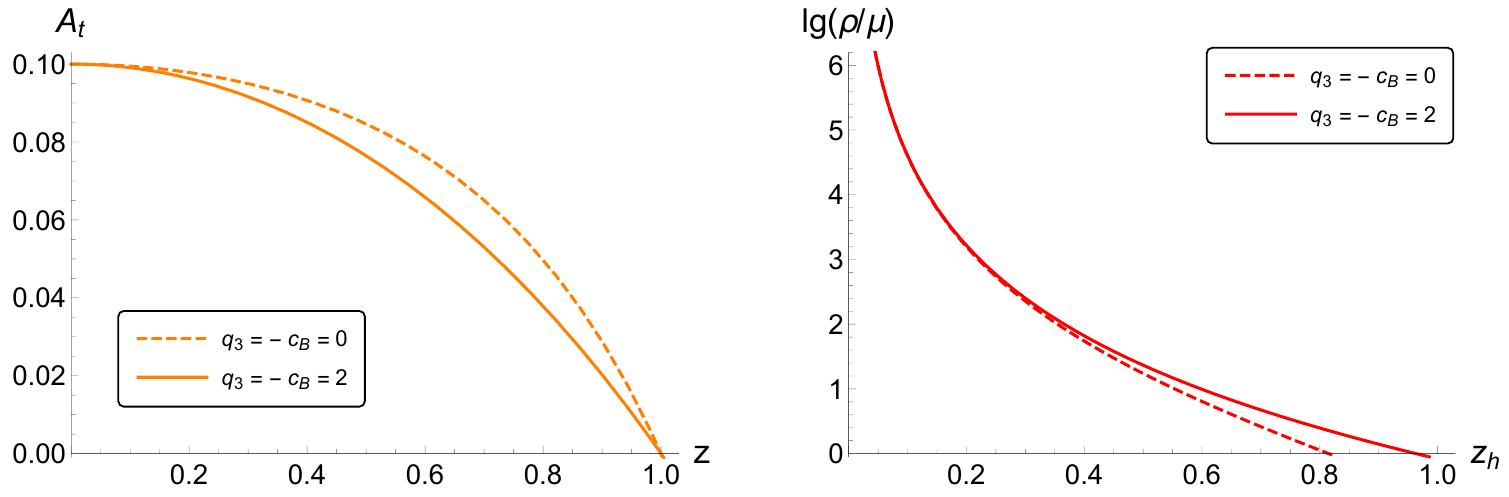} \\
  A \hspace{210pt} B
  \caption{Electric potential $A_t(z)$ (A) and density $\rho(z_h)/\mu$
    in logarithmic scale (B) for the ``heavy quarks'' case with (solid
    lines) and without (dashed lines) magnetic field, 
    $q_3 = - \, c_B = 0, \, 2$ (B); $R_{gg} = 1.16$; $[z]^{-1} =$ GeV.}
  \label{Fig:Atrho}
\end{figure}

Equation (\ref{eq:2.17}) with (\ref{eq:4.27}) gives
\begin{gather}
  A_t (z) = \mu \, \cfrac{e^{(2 R_{gg}+c_B(q_3-1))\frac{z^2}{2}} 
    - e^{(2 R_{gg}+c_B(q_3-1))\frac{z_h^2}{2}}}{1 - e^{(2
      R_{gg}+c_B(q_3-1))\frac{z_h^2}{2}}} 
  = \mu \left( 1 - \cfrac{1 - {e^{(2
          R_{gg}+c_B(q_3-1))\frac{z^2}{2}}}}{1 - {e^{(2
          R_{gg}+c_B(q_3-1))\frac{z_h^2}{2}}}}
  \right). \label{eq:4.28}
\end{gather}
For $q_3 = 1$ and $c_B = C$ the result (\ref{eq:4.28}) coincides with
the expressions (2.27) and (2.31) in \cite{He:2020fdi}:
\begin{gather}
  A_{t} (z) = \mu \left( 1 - \cfrac{1 - {e^{R_{gg} z^2}}}{1 -
      {e^{R_{gg} z_h^2}}} \right). \label{eq:4.29}
\end{gather}
Density is the coefficient in $A_t$ expansion:
\begin{gather}
  A_t(z) = \mu - \rho \,z^2 + \dots \ \Longrightarrow \
  \rho = - \, \cfrac{\mu \bigl(2 R_{gg} + c_B (q_3 - 1) \bigr)}{2
    \left(1 - e^{(2 R_{gg}+c_B(q_3-1))\frac{z_h^2}{2}}
    \right)}. \label{eq:4.30}
\end{gather}
The electric potential $A_t(z)$ and density $\rho(z_h)/\mu$ in
logarithmic scale are depicted in Fig.\ref{Fig:Atrho} in panels A and
B, respectively.

Equation (\ref{eq:4.22}) with (\ref{eq:4.27}) and (\ref{eq:4.28})
gives
\begin{gather}
  g(z) = e^{c_B z^2} \left[ 1 - \cfrac{I_1(z)}{I_1(z_h)}
    + \cfrac{\mu^2 \bigl(2 R_{gg} + c_B (q_3 - 1) \bigr) I_2(z)}{L^2
      \left(1 - e^{(2 R_{gg}+c_B(q_3-1))\frac{z_h^2}{2}}
      \right)^2} \left( 1 - \cfrac{I_1(z)}{I_1(z_h)} \,
      \cfrac{I_2(z_h)}{I_2(z)} \right) \right], \label{eq:4.31} \\
  I_1(z) = \int_0^z e^{\left(2R_{gg}-3c_B\right)\frac{\xi^2}{2}+3p\xi^4}
  \xi^{1+\frac{2}{\nu}} \, d \xi, \quad 
  I_2(z) = \int_0^z
  e^{\bigl(2R_{gg}+c_B\left(\frac{q_3}{2}-2\right)\bigr)\xi^2+3p\xi^4}
  \xi^{1+\frac{2}{\nu}} \, d \xi. \label{eq:4.32}
\end{gather}



\begin{figure}[t!]
  \centering
  \includegraphics[scale=1]{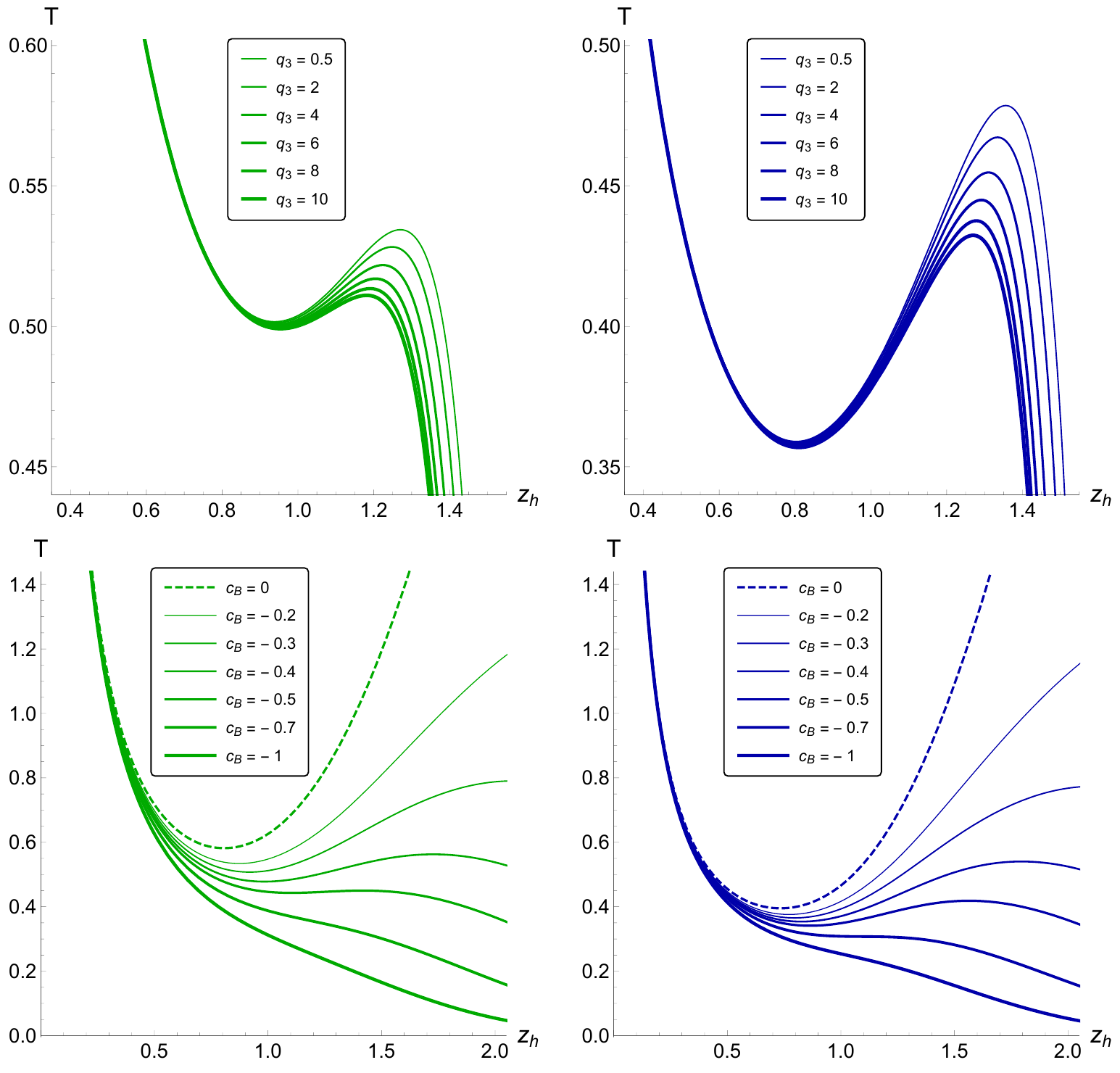} \\
  A \hspace{210pt} B \
  \caption{Temperature $T(z_h)$ for different $q_3$, $c_B = - \, 0.3$
    and $\mu =  \, 0.2$ (1-st line) and for different $c_B$, $q_3 = 1$
    and $\mu = \, 0$ (2-nd line); $\nu = 1$ (A) and  $\nu = 4.5$ (B);
    $R_{gg} = 1.16$, $p = 0.273$; $[T] = [z_h]^{-1} =$ GeV.}
  \label{Fig:Tzhq3cBmu0}
\end{figure}

We use the following formulas of temperature and entropy:
\begin{gather}
  T = \cfrac{\sqrt{{g_{tt}}' \, {g^{zz}}'}}{4 \pi} \, \Bigl|_{z=z_h}
  = \cfrac{\sqrt{{g_{00}}' \, {g^{44}}'}}{4 \pi} \,
  \Bigl|_{z=z_h}, \label{eq:4.33} \\
  s = \cfrac{\sqrt{g_{xx} \, g_{y_1y_1} \, g_{y_2y_2}}}{4} \,
  \Bigl|_{z=z_h} = \cfrac{\sqrt{g_{11} \, g_{22} \, g_{33}}}{4} \,
  \Bigl|_{z=z_h}. \label{eq:4.34}
\end{gather}
For the metric (\ref{eq:2.04}) and the warp factor (\ref{eq:4.23})
temperature and entropy can be written as:
\begin{gather*}
    T 
    = \cfrac{|g'|}{4 \pi} \, \Biggl|_{z=z_h}  = \Biggl|
    - \, \cfrac{e^{(2R_{gg}-c_B)\frac{z_h^2}{2}+3pz_h^4} \,
      z_h^{1+\frac{2}{\nu}}}{4 \pi \, I_1(z_h)} \, \times 
\end{gather*}
\begin{gather}
  \begin{split}
    & \times \left[ 
      1 - \cfrac{\mu^2 \bigl(2 R_{gg} + c_B (q_3 - 1) \bigr) 
        \left(e^{(2 R_{gg} + c_B (q_3 - 1))\frac{z_h^2}{2}}I_1(z_h) -
          I_2(z_h) \right)}{L^2 \left(1 
          - e^{(2R_{gg}+c_B(q_3-1))\frac{z_h^2}{2}}
        \right)^2} \right] \Biggr|, 
    \\
    s &= \cfrac{1}{4} \left( \cfrac{L}{z_h} \right)^{1+\frac{2}{\nu}}
    e^{-(2R_{gg}-c_B)\frac{z_h^2}{2}-3pz_h^4}. 
  \end{split}\label{eq:4.36} 
\end{gather}

\begin{figure}[t!]
  \centering
  \includegraphics[scale=1]{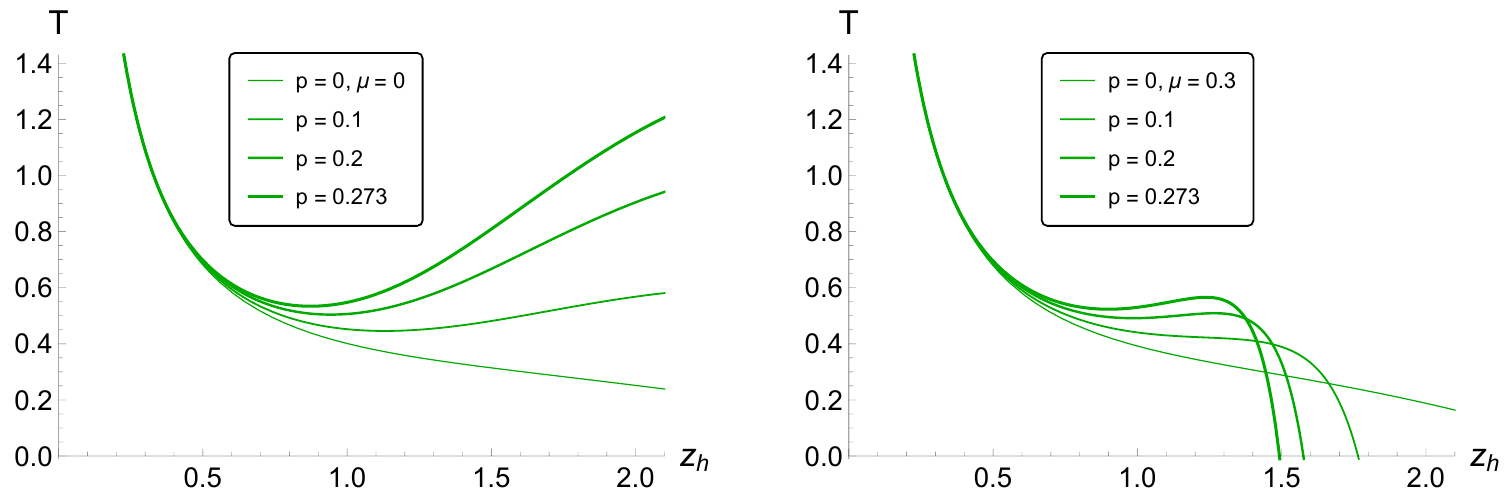} \\
  A \hspace{210pt} B \\
  \includegraphics[scale=1]{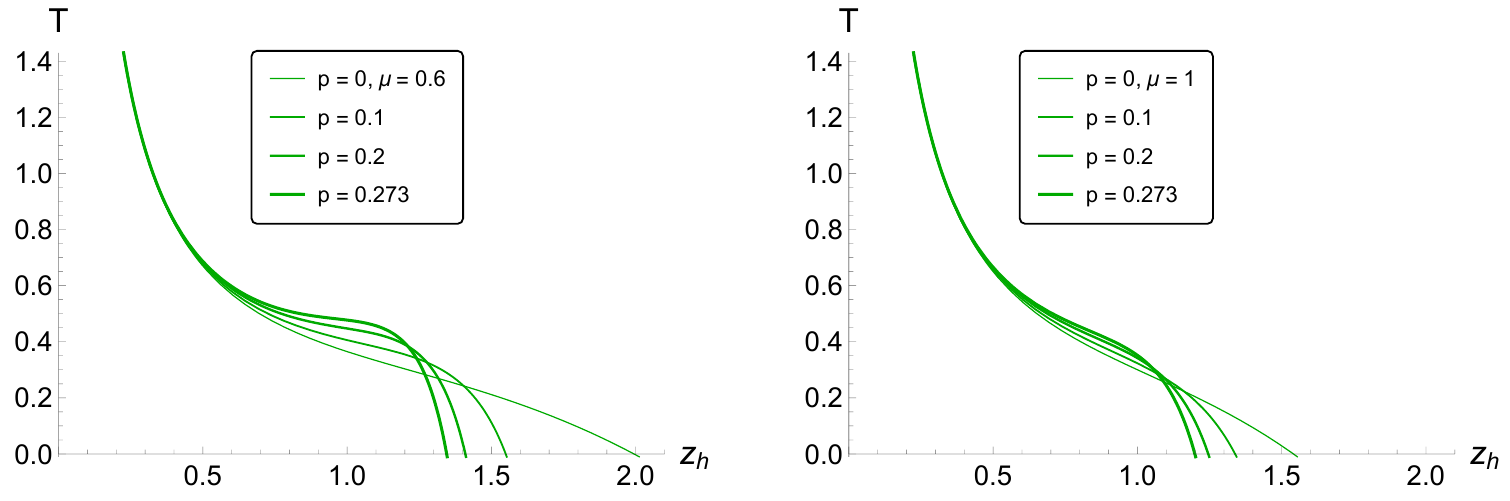} \\
  C \hspace{210pt} D
  \caption{Temperature $T(z_h)$ for different $p$, $\mu = 0$ (A), $\mu
    = 0.3$ (B), $\mu = 0.6$ (C), $\mu = 1$~(D); $\nu = 1$, $q_3 = - \,
    c_B = 0.2$, $R_{gg} = 1.16$; $[T] = [z_h]^{-1} =$ GeV.} 
  \label{Fig:Tzhpmunu1q302}
\end{figure}

\begin{figure}[t!]
  \centering
  \includegraphics[scale=0.99]{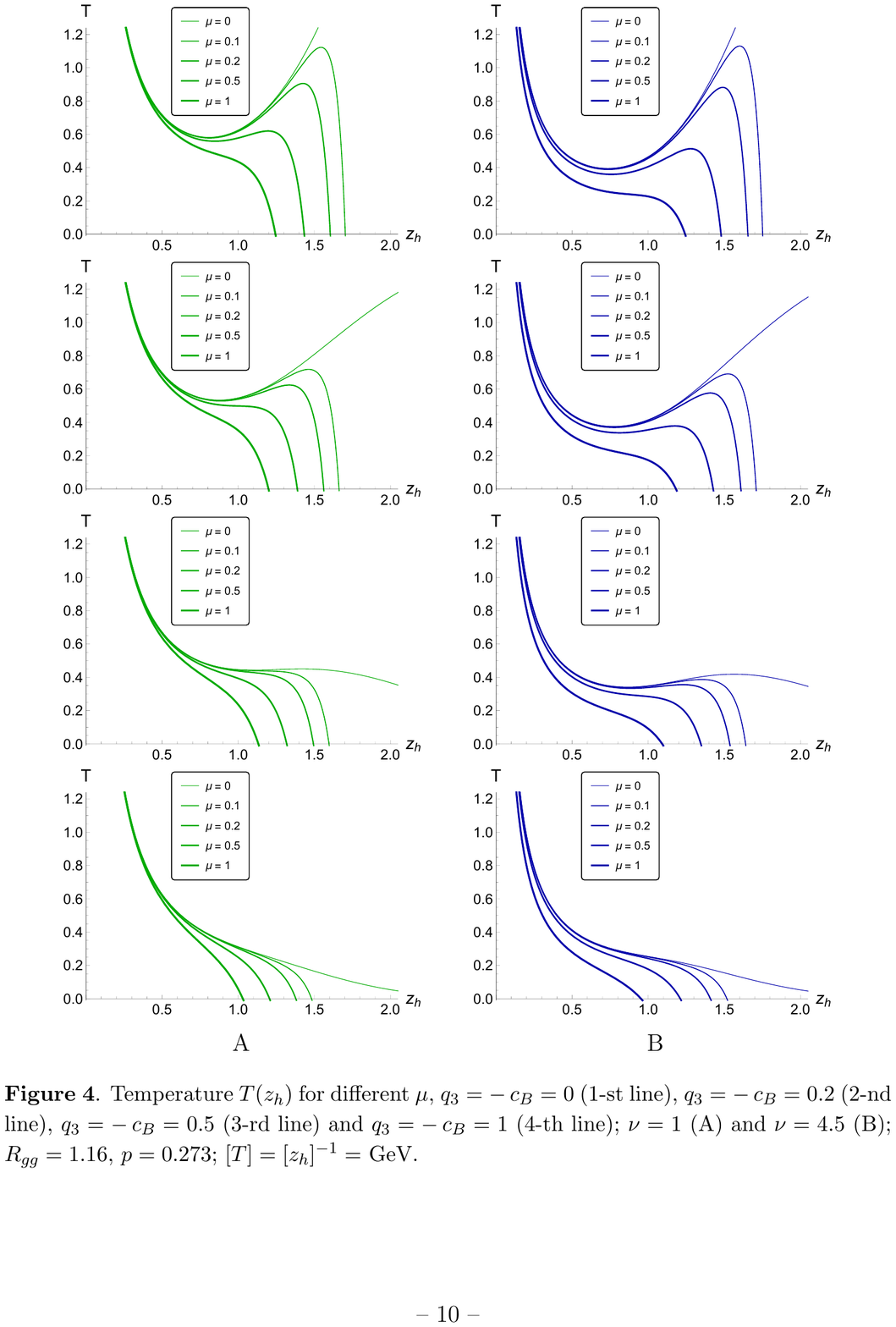} \\
  A \hspace{185pt} B
  \caption{Temperature $T(z_h)$ for different $\mu$, $q_3 = - \, c_B =
    0$ (1-st line), $q_3 = - \, c_B = 0.2$ (2-nd line),  $q_3 = - \,
    c_B = 0.5$ (3-rd line) and $q_3 = - \, c_B = 1$ (4-th line);  $\nu
    = 1$ (A) and $\nu = 4.5$~(B); $R_{gg} = 1.16$, $p = 0.273$; $[T] =
    [z_h]^{-1} =$ GeV.}
  \label{Fig:Tzhmu-q3-cB}
\end{figure}

In Fig.\ref{Fig:Tzhq3cBmu0} (first line) we see that magnetic
``charge'' $q_3$ affects the temperature function for the fixed
magnetic coefficient $c_B$ and non-zero chemical potential $\mu$
values independently from primary anisotropy. For zero chemical
potential temperature obviously has no dependence on $q_3$ at all, see
eq.~(\ref{eq:4.33}). At fixed $c_B$ increasing $q_3$ decreases  
the phase transition temperature and evenually makes it monotonic
(Fig.\ref{Fig:Tzhq3cBmu0}, second line) independently from primary
anisotropy as well. All this leads to the idea used in other works to
associate the magnetic ``charge'' with the magnetic coefficient $q_3 =
- \, c_B$.

\begin{figure}[t!]
  \centering
  \includegraphics[scale=1]{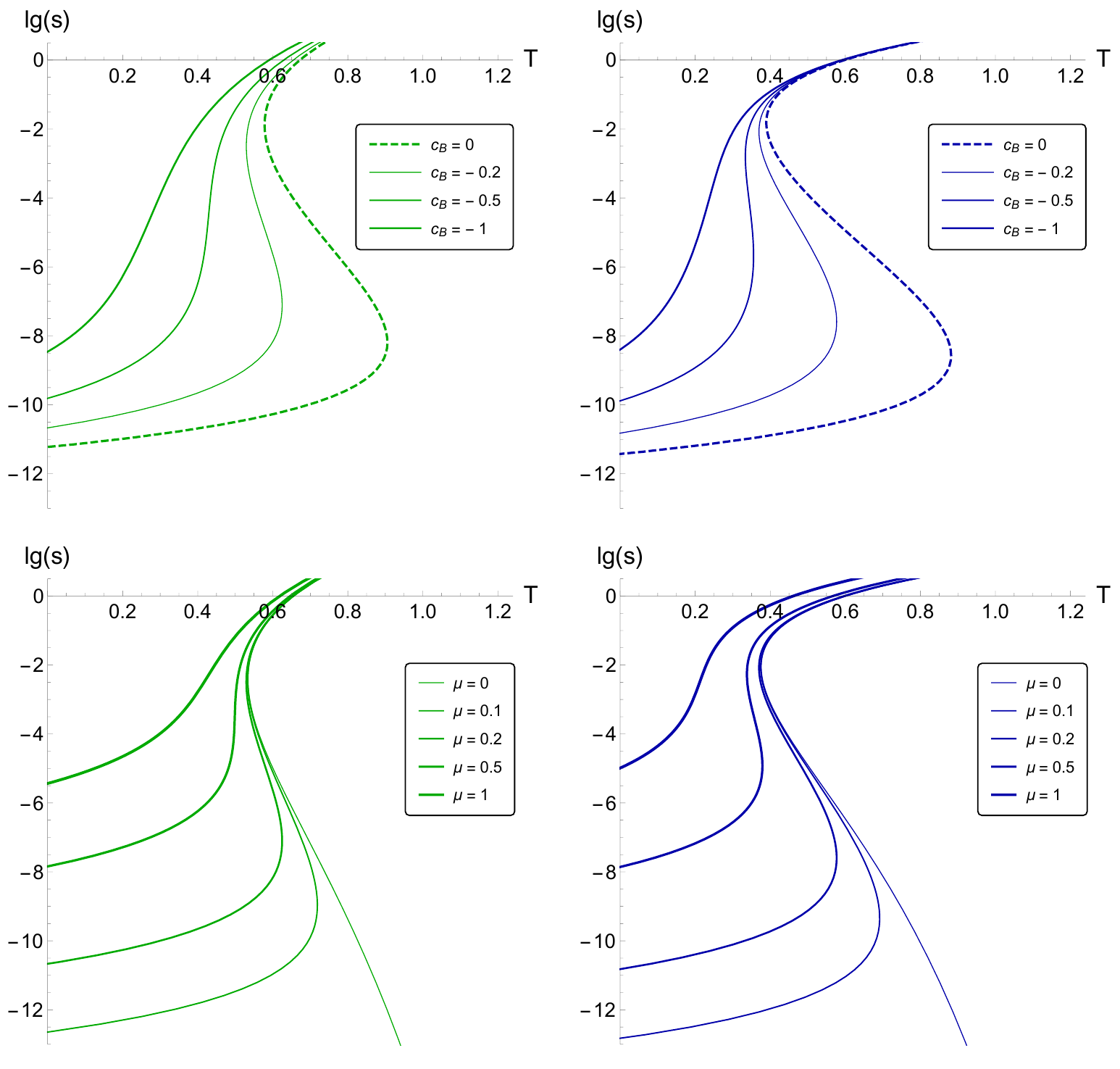} \\
  A \hspace{210pt} B
  \caption{Entropy $s(T)$ in logarithmic scale for different $q_3 = -
    \, c_B$, $\mu = 0.2$ (1-st line) and for different $\mu$, $q_3 = -
    \, c_B = 0.2$ (2-nd line); $\nu = 1$ (A) and $\nu = 4.5$ (B);
    $R_{gg} = 1.16$, $p = 0.273$; $[T]$ = GeV.}
  \label{Fig:sTcBmu}
\end{figure}

To understand the role of $pz^4$-term in the warp factor let's
consider the temperature behavior for $p = 0, \, 0.1, \, 0.2, \,
0.273$ (Fig.\ref{Fig:Tzhpmunu1q302}). Earlier in \cite{ARS-Heavy-2020}
we obtained a hypersensitivity of the background phase transition from
the magnetic field: for $c_B \sim - \, 0.01$ (for $\nu = 1$) it fully
degenerated. Non-zero $p$ makes the background phase transition in the 
magnetic field more stable, allowing it to achieve more realistic
values (in Fig.\ref{Fig:Tzhpmunu1q302} $q_3 = - \, c_B = 0.2$).

In Fig.\ref{Fig:Tzhmu-q3-cB} temperature function for different
chemical potential values and $p = 0.273$ in presented. Magnetic field
amplification makes three-valued behavior of $z_h(T)$ to become
monotonic and lowers the local minimum temperature. Entropy has usual
behavior (Fig.\ref{Fig:sTcBmu}) and the inverse magnetic catalysis can
be expected again. To make this sure let us consider a free energy
function

\begin{gather}
  F = - \int s \, d T = \int_{z_h}^{\infty} s \, T' dz.
  \label{eq:4.37}
\end{gather}
Note that for non-zero chemical potential we should integrate to
second horizon $z_{h_2}$ and for $\mu = 0$ we have $z_{h_2} =
\infty$.
 
\begin{figure}[t!]
  \centering
  \includegraphics[scale=0.8]{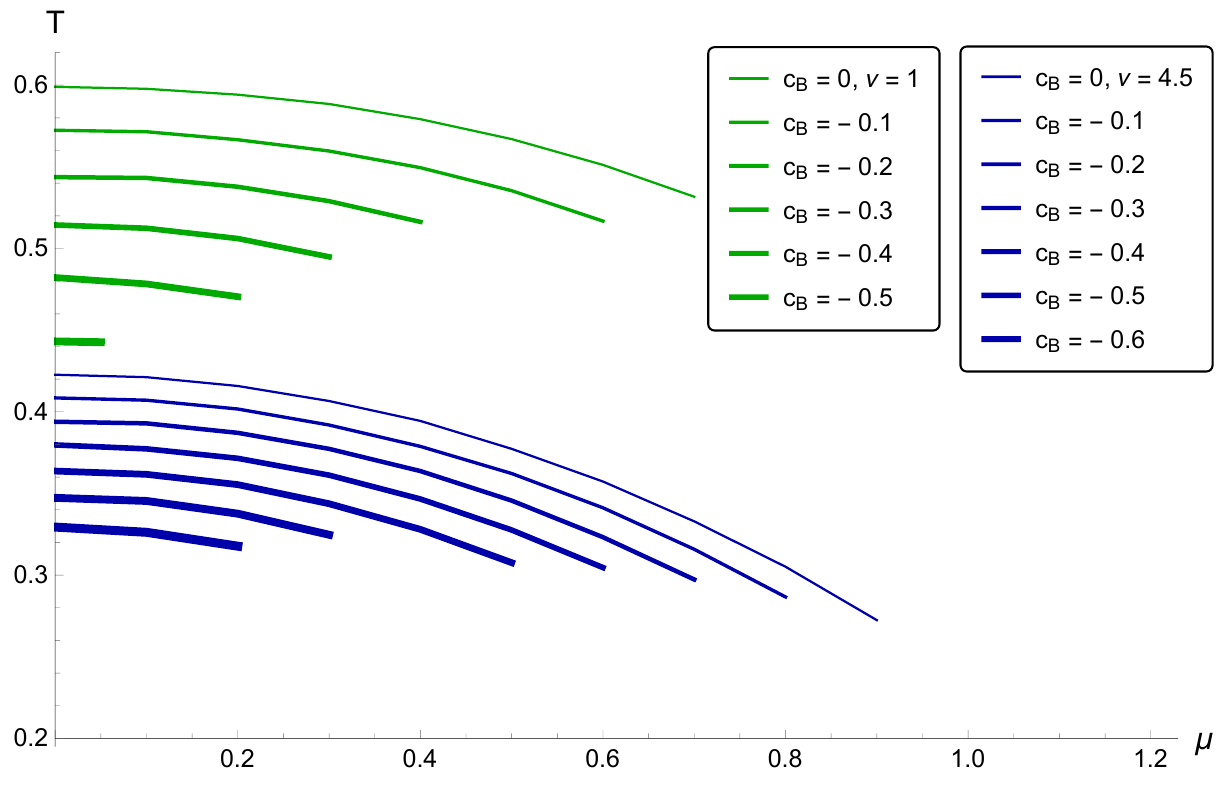} \\
  \caption{Phase diagram for $\nu = 1$ (green lines) and $\nu = 4.5$
    (blue lines) for $q_B = - \, c_B$; $R_{gg} = 1.16$, $p = 0.273$;
    $[T] = [\mu] =$ GeV.}
  \label{Fig:Tmu}
  \ \\ 
  \includegraphics[scale=1]{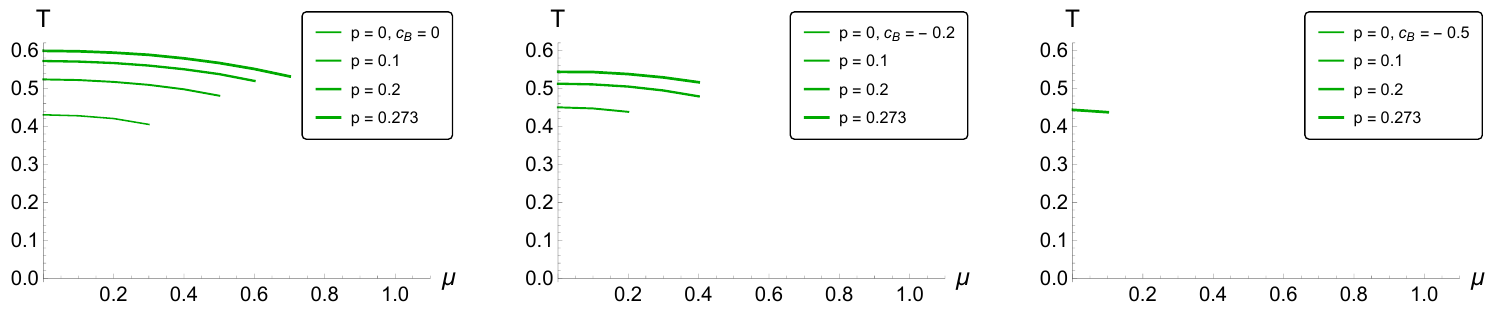} \\
  \quad A \hspace{130pt} B \hspace{130pt} C
  \caption{Phase diagram for different $p$, $q_B = - \, c_B = 0$ (A),
    $q_B = - \, c_B = 0.2$ (B), $q_B = - \, c_B = 0.5$ (C); $\nu = 1$,
    $R_{gg} = 1.16$, $p = 0.273$. }
  \label{Fig:Tmup}
\end{figure}

The result for the first order (background) black hole-black hole
(BB) phase transition or Hawking-Page-like (HP) phase transition
originating from the free energy is presented in Fig.\ref{Fig:Tmu} and
confirms the inverse magnetic catalysis effect. As to the $pz^4$-term
effect, in Fig.\ref{Fig:Tmup} we can clearly see that larger $p$-value
prevents the background phase transition degeneracy with the magnetic
field amplification.


Thus our choice of (\ref{eq:4.23}) as a warp factor to deform the
metric did not possess the MC phenomenon although could lead to good
results for IMC. In the next section we consider a new warp factor to
produce the proper result for MC phenomenon.



\subsection{Solution for ${\cA}(z) = - \, cz^2/4 \, - (p - c_B q_3)
  z^4$}

In search for the MC phenomenon we considered a corrected version of
our new warp factor. Let us remind that varying Lagrangian
(\ref{eq:2.01}) over the metric (\ref{eq:2.04}) we get the same
equations of motion (\ref{eq:2.16}--\ref{eq:2.22}) for different
fields. Using the new warp factor 
\begin{gather}
  \fb(z) = e^{2{\cA}(z)} = e^{ - \, c z^2/2 \, - \,2 (p-c_B \, q_3)
    z^4}, \label{eq:4.38}
\end{gather}
where $c = 4 R_{gg}/3$, $R_{gg} = 1.16$, $p = 0.273$, one can solve
system of EOMs (\ref{eq:2.16}--\ref{eq:2.22}) with the same boundary
conditions (\ref{eq:4.24}--\ref{eq:4.26}).

To possess the linear Regge trajectories for the meson mass spectra
in our model, in comparison with \cite{He:2020fdi}, we considered the
kinetic function (\ref{eq:4.27}) and the new warp factor
(\ref{eq:4.38}). Then, at $T = \mu = B = 0$ one can produce linear
mass spectrum $m_n^2 = 4 R_{gg} n$, in such a way that the parameter
$R_{gg}$ can be fitted by Regge spectra of meson, such as
$J/\psi$. Note also that the parameters $R_{gg}$ and $p$ can be fixed
for the zero magnetic field with $R_{gg} = 1.16$ and $p = 0.273$
\cite{He:2013qq,Li:2017tdz}.

\subsubsection{Blackening function}

\begin{figure}[t!]
  \centering
  \includegraphics[scale=1]{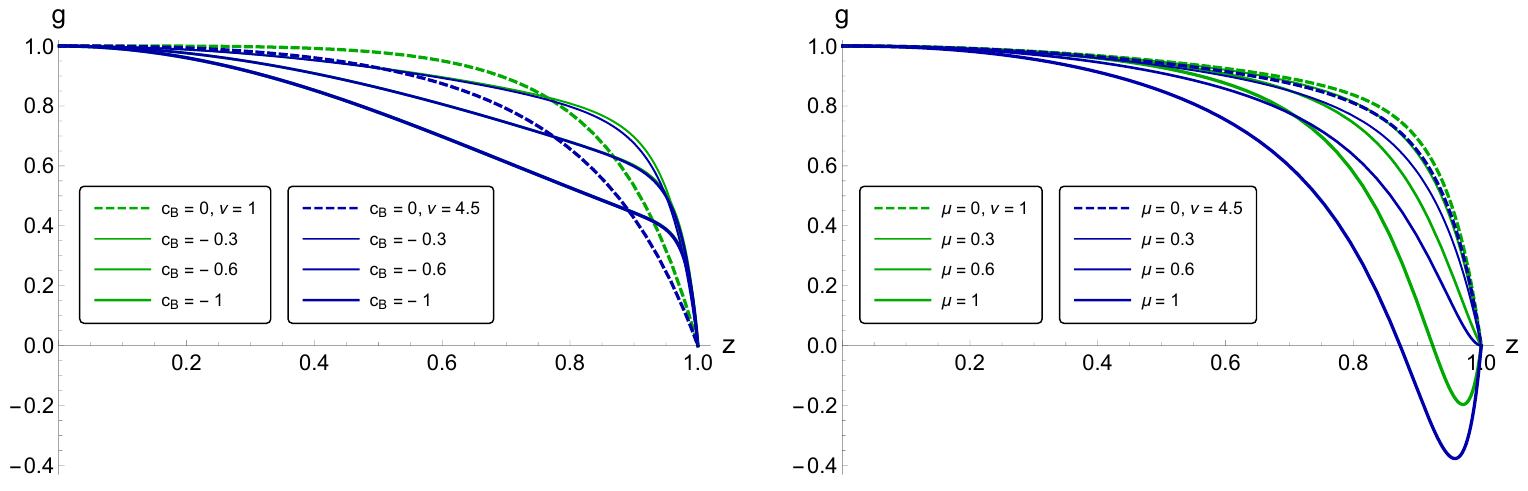} \\
  A \hspace{210pt} B
  \caption{The blackening function $g(z)$ for different $c_B$ and
    $\nu$, $\mu = 0$ (A) and for different $\mu$ and $\nu$, $c_B = -
    \, 0.2$ (B); $\nu = 1$ (green lines) and $\nu = 4.5$ (blue lines);
    $R_{gg} = 1.16$, $p = 0.273$, $q_3=5$, $z_h=1$; $[z]^{-1} =$ GeV.}
  \label{Fig:gvszmucB}
\end{figure}

For the corrected factor (\ref{eq:4.38}) the EOM on the gauge field
$A_t(z)$ is the same as before and has the same solution
(\ref{eq:4.28}). Therefore, the equation (\ref{eq:4.22}) with
(\ref{eq:4.27}), (\ref{eq:4.28}) and the corrected warp factor
(\ref{eq:4.38}) gives
\begin{gather}
  g(z) = e^{c_B z^2} \left[ 1 - \cfrac{\Tilde{I}_1(z)}{\Tilde{I}_1(z_h)}
    + \cfrac{\mu^2 \bigl(2 R_{gg} + c_B (q_3 - 1) \bigr)
      \Tilde{I}_2(z)}{L^2 \left(1 - e^{(2 R_{gg}+c_B(q_3-1))\frac{z_h^2}{2}}
      \right)^2} \left( 1 - \cfrac{\Tilde{I}_1(z)}{\Tilde{I}_1(z_h)} \,
      \cfrac{\Tilde{I}_2(z_h)}{\Tilde{I}_2(z)} \right)
  \right], \label{eq:4.42} \\
  \Tilde{I}_1(z) = \int_0^z
  e^{\left(2R_{gg}-3c_B\right)\frac{\xi^2}{2}+3 (p-c_B \, q_3) \xi^4}
  \xi^{1+\frac{2}{\nu}} \, d \xi, \qquad \ \label{eq:4.43-1} \\
  \Tilde{I}_2(z) = \int_0^z
  e^{\bigl(2R_{gg}+c_B\left(\frac{q_3}{2}-2\right)\bigr)\xi^2+3 (p-c_B
    \, q_3) \xi^4} \xi^{1+\frac{2}{\nu}} \, d \xi. \label{eq:4.43} 
\end{gather}
The behavior of the blackening function $g$ in terms of the
holographic coordinate $z$ for different values of the magnetic
coefficient $c_B$ and different primary anisotropy background,
i.e. $\nu = 1$ (green lines) and $\nu = 4.5$ (blue lines), normalized
to the horizon $z_h = 1$, is depicted in
Fig.\ref{Fig:gvszmucB}.A. Here blackening function is monotonic, and
larger values of magnetic coefficient $c_B$ correspond to lower $g$
values both in isotropic and anisotropic background. But comparing
isotropic and anisotropic backgrounds we see that at $0 < |c_B| < 0.3$
the blackening function gets lower values for $\nu = 4.5$ than for
$\nu = 1$, while at $0.3 < |c_B| < 1$ it is not sensible to changes in
primary anisotropies.

The effect of chemical potential on the blackening function for
different primary anisotropies of the background is demonstrated in
Fig.\ref{Fig:gvszmucB}.B. Larger $\mu$ decreases the blackening
function value in both isotropic and anisotropic background
cases. But, for the fixed chemical potential the blackening function
value is smaller in the background with larger primary
anisotropy. Note also that for large chemical potential values one has
to deal with the second black hole horizon.

\subsubsection{Scalar field}

\begin{figure}[t!]
  \centering
  \includegraphics[scale=1]{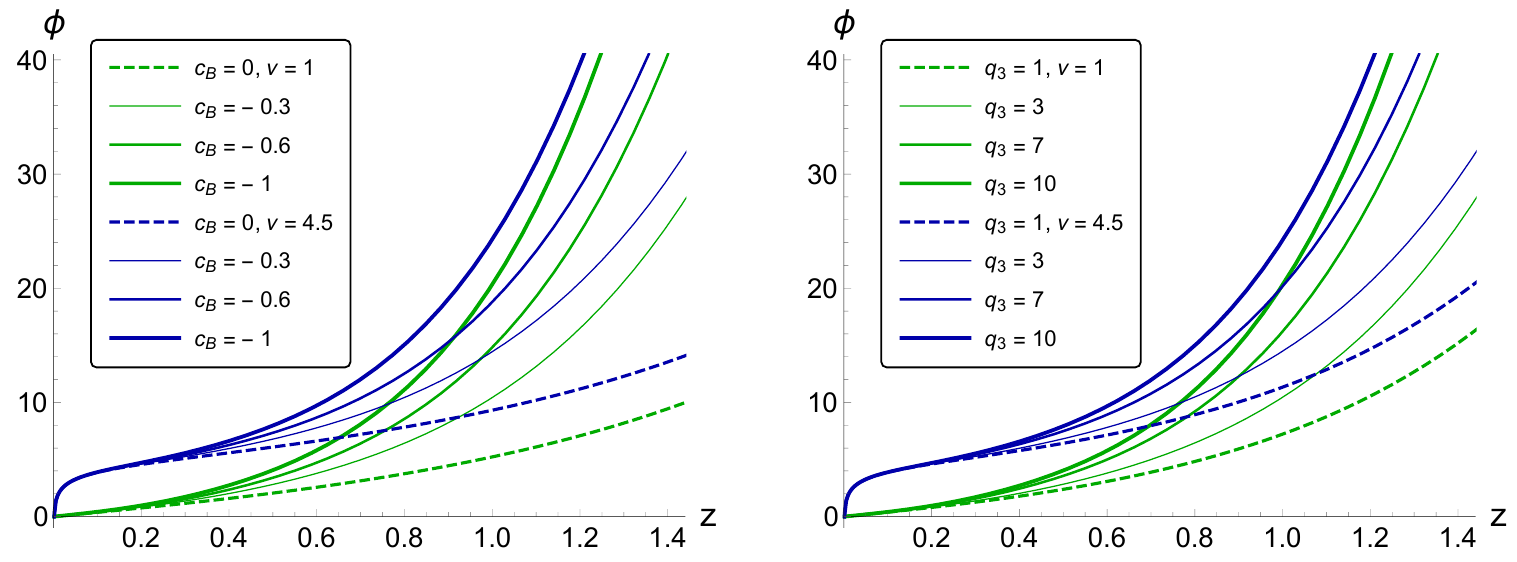} \\
  A \hspace{210pt} B
  \caption{The dilaton field $\phi(z)$ for different $c_B$, $q_3 = 5$
    (A) and different $q_3$, $c_B = - \, 0.5$ (B); $\nu = 1$ (green
    lines) and $\nu = 4.5$ (blue lines); $R_{gg} = 1.16$, $p = 0.273$,
    $z_0 = 0.001$; $[z]^{-1} =$ GeV.}
  \label{Fig:phivszcBq3}
\end{figure}

The scalar, i.e. dilaton field $\phi$ can be derived from equation
(\ref{eq:2.19})\footnote{This is a generalization of the method
  \cite{Gubser:2008ny, DeWolfe:2013cua} of reconstructing the dilaton
  potential, see also \cite{Li:2011hp, Cai:2012xh, Cai:2012eh}.} with
the boundary condition (\ref{eq:4.26}) imposed

\begin{gather}
  \begin{split}
    \phi (z) &= \int_{z_0}^z \Bigl[
    - \, 4 + \frac{2}{3} \, \nu \Biggl(
    6 + 3 \, (c_B \, (2 - 3 \nu) + 6 R_{gg} \nu) \, \xi^2
    + \bigl(- \, 3 \, c_B(c_B + 60 \, q_3) \, + \\
    &+ 4 (45 \, p + R_{gg}^2) \bigr) \, \nu \xi^4
    + 48 R_{gg} (p - c_B \, q_3) \, \nu \xi^6 
    + 144 \, (p-c_B \, q_3)^2 \, \nu \xi^8 \Bigr)
    \Biggr]^{\frac{1}{2}} \, \frac{d\xi}{\nu \xi}.
 \end{split}\label{eq:4.80}
\end{gather}

Expanding the integrand of the dilaton field we have $\phi(z) \sim
\int_{z_0}^{z} \sqrt{\nu-1} \, d\xi/\xi$. Therefore, the dilaton field
has no divergency at $z_0=0$ on the primary isotropic background $\nu
= 1$, while on anisotropic background a logarithmic divergency exists.
It is important to note that we generalize the boundary condition for
the dilaton field as $\phi(z_0) = 0$ \cite{ARS-Light-2020,He:2010ye},
where $z_0$ can be  some function of $z_h$. The fact is that the
scalar field boundary conditions can affect the temperature dependence
of the string tension, i.e. the coefficient of the linear term of the
Cornell potential. The string tension should decrease with the
temperature growth and become zero at the confinement/deconfinement
phase transition \cite{Digal:2005ht,Cardoso:2011hh,Bicudo:2010hg}. To
preserve this feature and also avoid divergences in anisotropic
backgrounds we considered the dilaton boundary condition as
$\phi(z_0) = 0$. For special cases one can consider $z_0 = 0$
\cite{Li:2017tdz} and $z_0 = z_h$ \cite{AR-2018}.

Fig.\ref{Fig:phivszcBq3} shows that the scalar field is a
monotonically increasing function of the holographic coordinate $z$
both in primary isotropic and anisotropic cases, i.e. for $\nu = 1$
and $\nu = 4.5$, but larger primary anisotropy shifts the dilaton
curve up to larger $\phi$-values. Larger absolute value of the
magnetic coefficient $c_B$ and larger magnetic charge $q_3$ make
$\phi(z)$ to grow faster (Fig.\ref{Fig:phivszcBq3}.A and B
respectively).

\subsubsection{Coupling function $f_3$}

\begin{figure}[t!]
  \centering
  \includegraphics[scale=1]{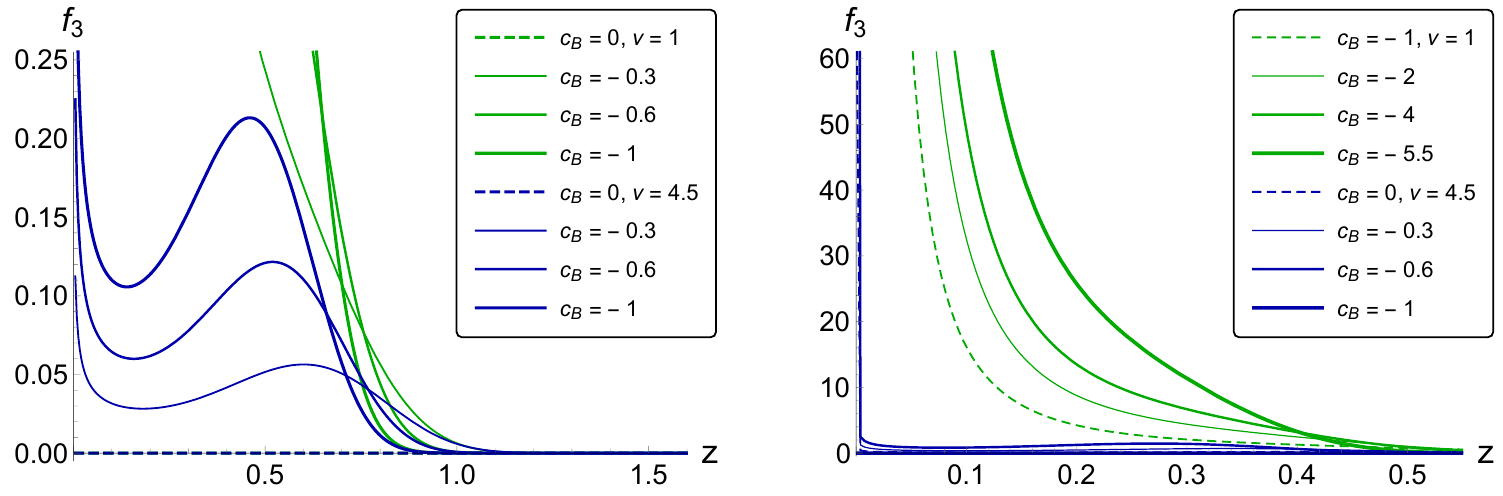} \\
  A \hspace{210pt} B
  \caption{The coupling function $f_3(z)$ for different $c_B$, $\mu =
    0$ (A) and for different $c_B$, $\mu = 0.2$ (B); $\nu = 1$ (green
    lines) and $\nu = 4.5$ (blue lines); $z_h = 1.5$, $R_{gg} = 1.16$,
    $p = 0.273$, $q_3=5$; $[z]^{-1} =$ GeV.}
  \label{Fig:f3vszcBdiff}
\end{figure}

The function $f_3$ that describes the coupling between the third
Maxwell field $F_3$ and the dilaton field $\phi$ is still calculated
by the expression (\ref{eq:4.21}). The detailed formula obtained by
substituting the blackening function, it's derivative and the
corrected warp factor (\ref{eq:4.38}) is presented in Appendix
\ref{blackf}. For zero magnetic coefficient $c_B = 0$ the coupling
function $f_3$ obviously equals zero. It's behavior depending on the
holographic direction $z$ is plotted in Fig.\ref{Fig:f3vszcBdiff}. We
see that for zero chemical potential $\mu = 0$ it is positive and
preserves the NEC both in isotropic $\nu = 1$ and primary anisotropic
$\nu = 4.5$ backgrounds (Fig.\ref{Fig:f3vszcBdiff}.A). However in the
isotropic background we see the decreasing monotonic behavior of
$f_3(z)$, while in the anisotropic background it demonstrates a
multivalued behavior with a local minimum and a local maximum. Note
also that for larger magnetic coefficient (larger
$c_B$ absolute values) $f_3$ is positive not everywhere beyond the
fixed horizon $z_h$, therefore we need to choose
appropriate parameters in our theory and in particular the correct
value for the second horizon to have positive value for coupling
function $f_3$ (see the Fig.\ref{Fig:f3vszcBdiff}.B for the second
horizon $z_{h_2} = 0.55$ for $\nu = 1$ and $z_{h_2} = 0.45$ for $\nu =
4.5$).

\begin{figure}[t!]
  \centering
  \includegraphics[scale=1]{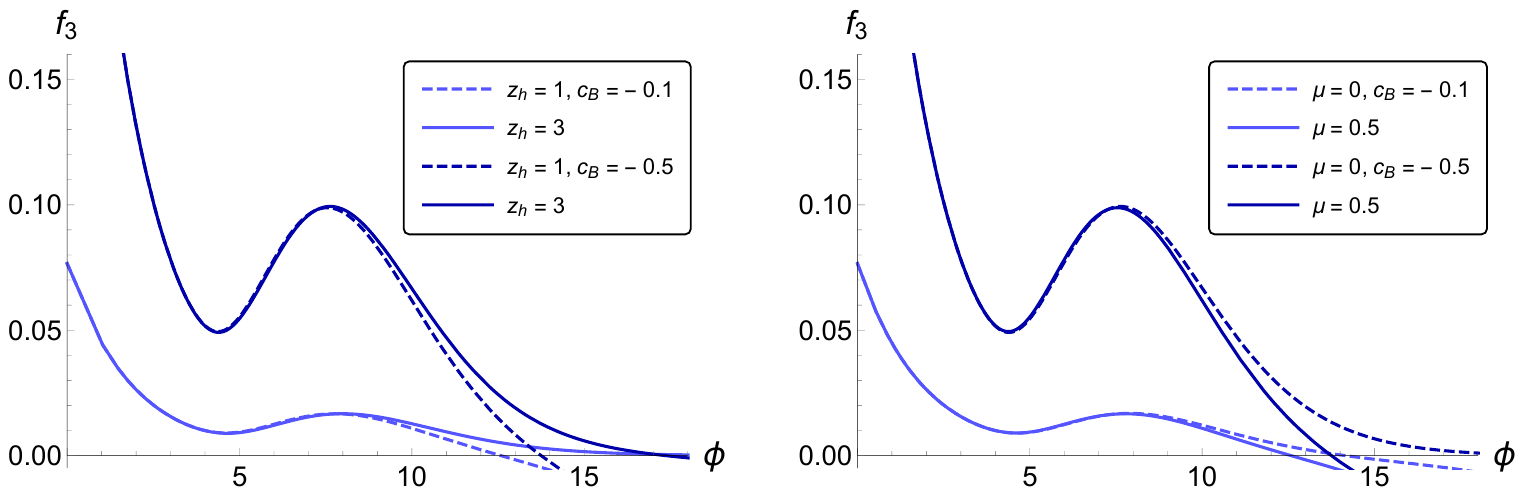} \\
  A \hspace{210pt} B
  \caption{The coupling function $f_3(\phi)$ different $z_h$ and
    $c_B$, $\mu = 0.5$ (A), for different $\mu$ and $c_B$, $z_h = 1$
    (B); $\nu = 4.5$, $R_{gg} = 1.16$, $p = 0.273$, $q_3 = 5$, $z_0 =
    0.001$; $[f_3] =$ GeV$^{-2}$.}
  \label{Fig:f3vsphi}
\end{figure}

According to the boundary condition $g(z_h) = 0$ expression
(\ref{eq:4.21}) is simplified as

\begin{gather}
  f_3(z_h) = 2 \left( \cfrac{L}{z_h} \right)^{\frac{2}{\nu}}
   \ \cfrac{c_B z_h}{q_3^2} \ \fb(z_h) \
  g'(z_h) > 0. \label{eq:3.12}
\end{gather}
At the first horizon (the one with smaller $z_h$ value, that really
matters) the blackening function is decreasing, so $g'(z_h) < 0$ and
temperature $T(z_h) = |g'(z_h)/4\pi|$ is positive. If we also take
$c_B < 0$, their product is positive, all the other multipliers in
(\ref{eq:3.12}) are positive as well, therefore $f_3(z_h) > 0$ for any
negative $c_B$ in the $z$ interval we need for $0 < z < z_{h_{min}}$,
where $z_{h_{min}}$ is not the fixed horizon, but the horizon with
smaller value, i.e. $T(z_{h_{min}}) = 0$. For zero chemical potential
$z_{h_{min}} = z_h$ and for $\mu = 1$, for example, $z_{h_{min}} =
z_{h_2}$ (Fig.\ref{Fig:gvszmucB}.B).

In Fig.\ref{Fig:f3vsphi} the third coupling function $f_3$ in terms of
dilaton field $\phi$ is displayed. It demonstrates a nonmonotonic
behavior, quite sensible to the magnetic field presence -- larger
$c_B$ absolute value leads to larger $f_3$. Neither fixed horizon
(Fig.\ref{Fig:f3vsphi}.A), nor chemical potential
(Fig.\ref{Fig:f3vsphi}.B) have no significant effect.

\subsubsection{Coupling function $f_1$}

We also need to check the NEC for the function $f_1$ in our model. It
describes coupling between the second Maxwell field $F_1$ and the
dilaton field $\phi$:
\begin{gather}
  f_1 = - \left( \cfrac{z}{L} \right)^{1-\frac{4}{\nu}} 
  \cfrac{e^{c_Bz^2} \, \fb \, (\nu -1) }{L \, q_1^2 \, \nu}  \left[ 
    2 g' + 3 g \left(
      \cfrac{\fb'}{\fb} - \cfrac{4 \left( \nu + 1 \right)}{3 \nu z}
      + \cfrac{2 c_B z}{3}
    \right) \right]. \label{eq:2.29}
\end{gather}
The exact formula obtained by substituting the blackening function,
its derivative and the new warp factor is presented in Appendix
\ref{blackf}. It is clearly seen that the coupling function $f_1$ is
zero for $\nu = 1$, as according to the holographic dictionary the
second Maxwell field $F_1$ serves for primary anisotropy of the
background.

\begin{figure}[b!]
  \centering
  \includegraphics[scale=1]{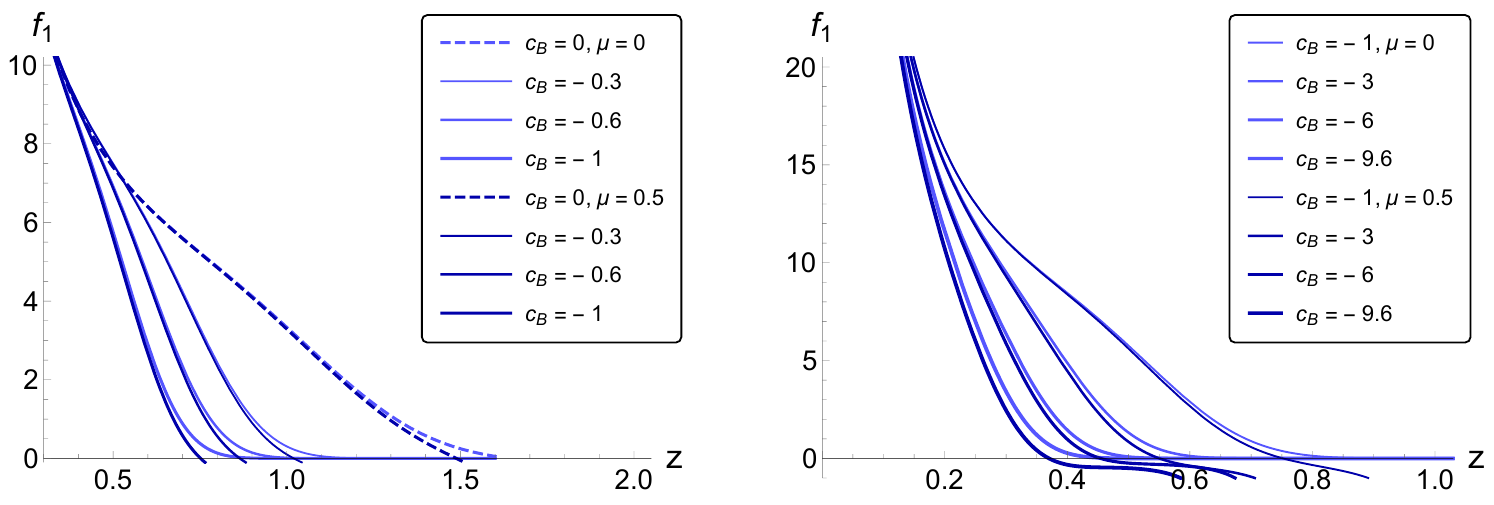} \\
  A \hspace{210pt} B
  \caption{The coupling function $f_1(z)$ for different $c_B \ge - \,
    1$, $\mu = 0$ (light blue) and $\mu = 0.5$ (dark blue),
    $z_{h_{min}} \ge 0.73$ (A); for different $c_B \le - \, 1$, $\mu =
    0.5$, $z_{h_{min}} \ge 0.39$ (B); $\nu = 4.5$, $z_h = 1.5$,
    $R_{gg} = 1.16$, $p = 0.273$, $q_3 = 5$, $q_1 = 1$; $[z]^{-1}
    =$~GeV.}
  \label{Fig:f1vszcBdiff}
\end{figure}

Fig.\ref{Fig:f1vszcBdiff} shows the coupling function $f_1$ in terms
of holographic coordinate $z$. It gets positive value for zero
chemical potential (Fig.\ref{Fig:f1vszcBdiff}.A, light blue), so the
NEC is fulfilled in our model. For perceptible chemical potential
value $\mu = 0.5$ (Fig.\ref{Fig:f1vszcBdiff}.A, dark blue) the
coupling function $f_1$ can be not positive for $z < z_h$ in magnetic
field strong enough. But the second horizon shift under the first
(fixed) one must be taken in account again. For $z < z_{h_2}$ the
coupling function $f_1$ stays positive similar to the coupling
function $f_3$, thus NEC is respected (Fig.\ref{Fig:f1vszcBdiff}.B,
the second horizon $z_{h_{min}} = z_{h_2} = 0.39$). For $f_1$ it can
be proven algebraically, like it was done for it in
\cite{ARS-Heavy-2020} and for $f_3$ in previous section, that it
doesn't break the NEC either. According to the boundary condition
$g(z_h) = 0$ expression for $f_1$ can be simplified as

\begin{gather}
  f_1(z_h) = -\,2 \left( \cfrac{z_h}{L} \right)^{1-\frac{4}{\nu}}
   \cfrac{\nu - 1}{L\, q_1^2 \nu } \ e^{c_B z_h^2} \ \fb(z_h)  \,
   g'(z_h). \label{eq:3.13}
\end{gather}
At the first horizon $z_h = z_{h_{min}}$ the blackening function is
decreasing, so $g'(z_h) < 0$ and temperature $T(z_h)=|g'(z_h)/4\pi|$
is positive. As $f_1$ has sense for $\nu > 1$ only, all the
multipliers in (\ref{eq:3.13}) are positive, so their product is
positive, therefore $f_1(z_h) > 0$ for any $z_h$ from the interval $0<
z_h < z_{h_{min}}$, where $z_{h_{min}}$ is the actual horizon,
i.e. $T(z_{h_{min}}) = 0$.

\begin{figure}[t!]
  \centering
  \includegraphics[scale=1]{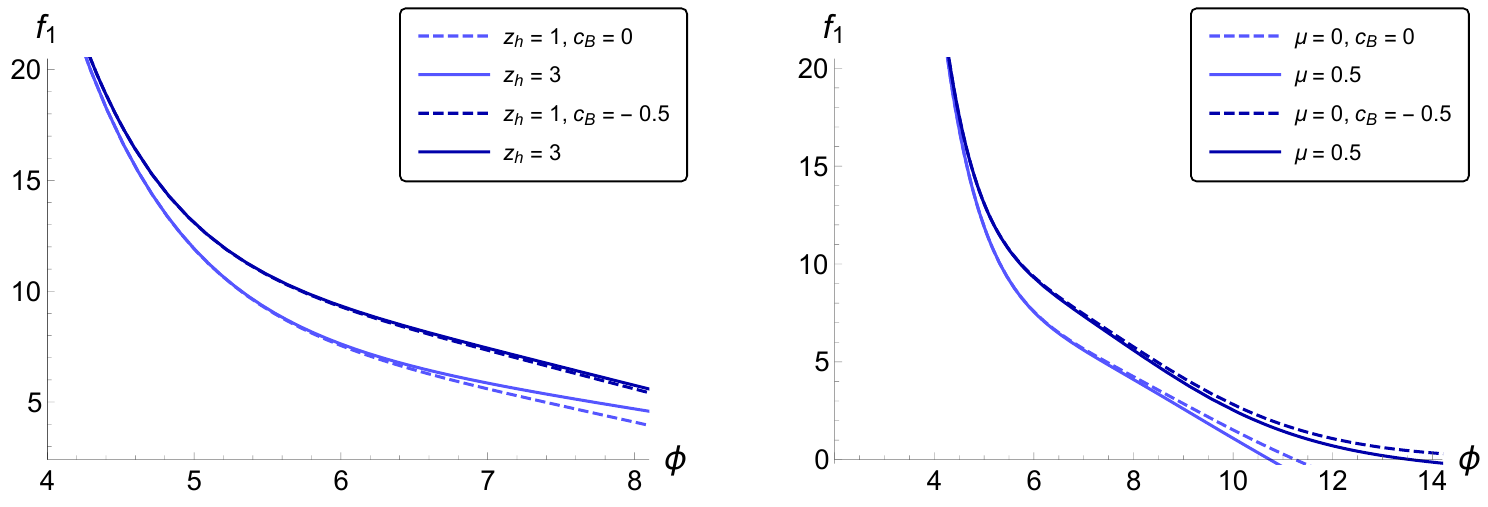} \\
  A \hspace{210pt} B
  \caption{The coupling function $f_1(\phi)$ for different $z_h$ and
    $c_B$, $\mu = 0.5$ (A), for different $\mu$ and $c_B$ and $z_h =
    1$; $\nu = 4.5$, $R_{gg} = 1.16$, $p = 0.273$, $q_3 = 5$, $z_0 =
    0.001$.}
  \label{Fig:f1vsphi}
\end{figure}

For the completeness of the results presented the second coupling
function $f_1(\phi)$ is also plotted (Fig.\ref{Fig:f1vsphi}). It turns
out to be insensitive to the magnetic field for small $\phi$ values
and almost insensitive to the fixed horizon value even for large
$\phi$ (Fig.\ref{Fig:f1vsphi}.A). Besides, the magnetic field shifts
$f_1$ up to larger values, while the chemical potential, on the
contrary, downs to smaller values (Fig.\ref{Fig:f1vsphi}.B).

\subsubsection{Scalar potential}

\begin{figure}[b!]
  \centering
  \includegraphics[scale=1]{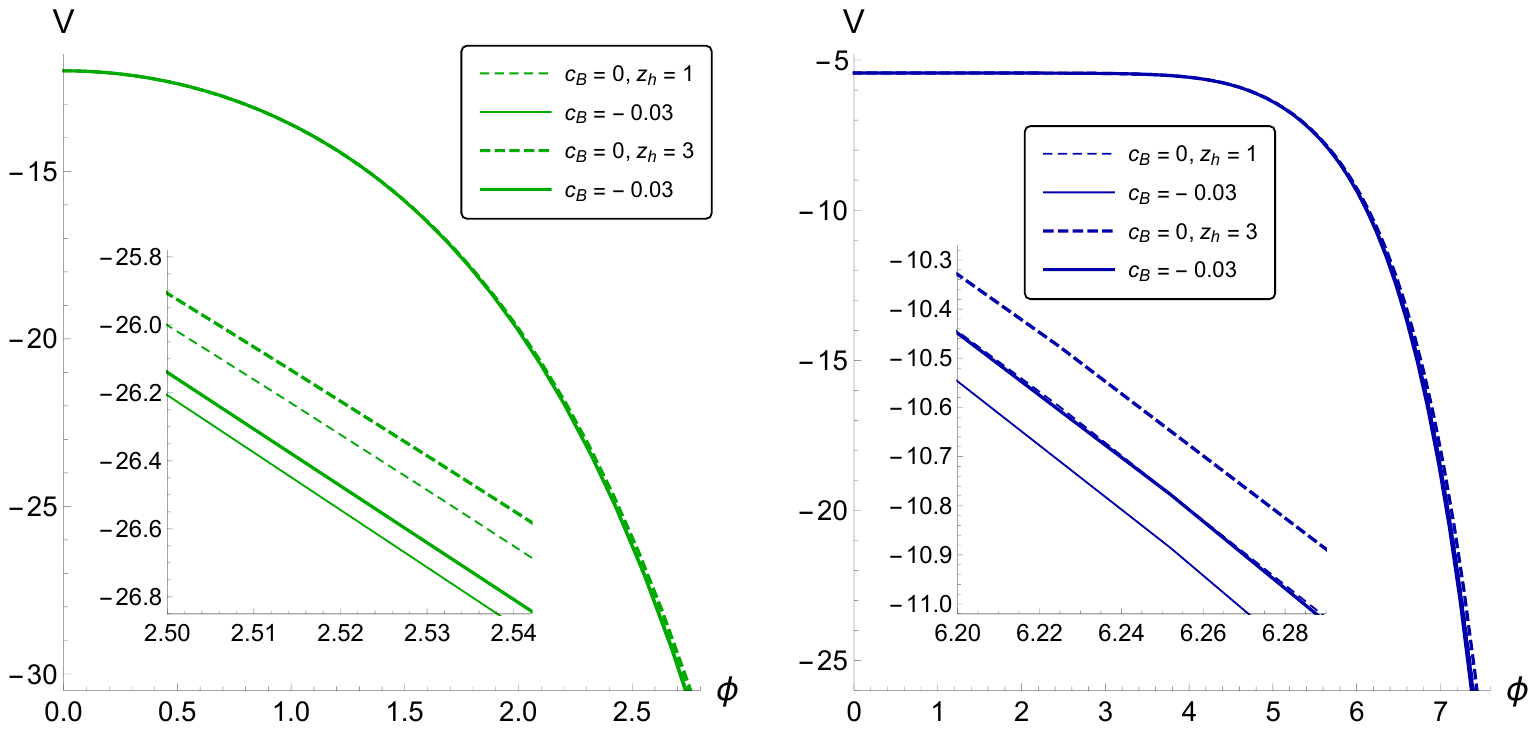} \\
  A \hspace{210pt} B
  \caption{The scalar potential $V(\phi)$ for different $c_B$ and
    $z_h$, $\nu = 1$ (A) and $\nu = 4.5$ (B); $R_{gg} = 1.16$, $p =
    0.273$, $q_3 = 5$, $\mu = 0.5$, $z_0 = 0.001$; $[V] =$ GeV$^2$.}
  \label{Fig:vvsphinu145}
\end{figure}

The equation of motion for the scalar potential $V$ as a function of
holographic coordinate $z$ can be obtained from the equation
(\ref{eq:2.22}):
\begin{gather}
  \begin{split}
    V(z) = - \, \cfrac{3}{2 L^2 \nu^2 \fb} 
    &\left[ \Biggl\{ 
      \cfrac{\fb''}{\fb} \ \nu^2 z^2
      + \cfrac{(\fb')^2}{2 \fb^2} \ \nu^2 z^2
      - \cfrac{\fb'}{\fb} \ \bigl( 3 (1 + \nu) - 2 c_B \nu z^2 \bigr)
      \nu z \right. + \\
    &\ \ + \cfrac{2}{3} \ \bigl( 2 (1 + \nu) (1 + 2 \nu) - (3 + 2
    \nu) c_B \nu z^2 + c_B^2 \nu^2 z^4 \bigr) \Biggr\} \, g \ + \\
    &\ \ + \left. \left\{ \cfrac{3 \fb'}{2 \fb} \, \nu z - \cfrac{4 +
          5 \nu}{3} + c_B \nu z^2 \right\} \nu z g'
      + \cfrac{\nu^2 z^2}{3} \, g'' \right].
  \end{split}\label{eq:4.81}
\end{gather} 
The exact formula obtained by substituting the blackening function,
its first and second derivatives and the new corrected warp factor is
presented in Appendix \ref{blackf}.

The scalar potential $V$ as a function of dilaton field $\phi$ is
described by equations (\ref{eq:4.80}) and (\ref{eq:4.81}) and is
depicted in  Fig.\ref{Fig:vvsphinu145}. Within the interval $0 \le z
\le z_h$, that we are interested in, $V(\phi)$ is stable and has no
crucial dependence either on the horizon nor on the magnetic
field. More focus in the scale  allows to see that the magnetic field
increases the absolute value of the scalar potential, while $z_h$
growth decreases it. Primary anisotropy (Fig.\ref{Fig:vvsphinu145}.B
for $\nu = 4.5$) shifts the scalar potential to larger absolute values
(Fig.\ref{Fig:vvsphinu145}.A for $\nu = 1$), causes a constant region
to appear at small $\phi$ and then makes $V(\phi)$ to decrease faster.

\subsection{Thermodynamics for ${\cA}(z) = - \, cz^2/4 \, - (p-c_B
  q_3) z^4$}

\subsubsection{Temperature and entropy}

Using the metric (\ref{eq:2.04}) and the warp factor $\fb(z) = e^{ -
  \, c z^2/2 \, - \,2 (p-c_B \, q_3) z^4}$ (\ref{eq:4.38}) one can
obtain the temperature and entropy from equations (\ref{eq:4.33}) and
(\ref{eq:4.34}) respectively:
\begin{gather}
  \begin{split}
    T &= \cfrac{|g'|}{4 \pi} \, \Biggl|_{z=z_h} = \left|
     - \, \cfrac{e^{(2R_{gg}-c_B)\frac{z_h^2}{2}+3 (p-c_B \, q_3)
         z_h^4} \, 
      z_h^{1+\frac{2}{\nu}}}{4 \pi \, \Tilde{I}_1(z_h)} \right. \times \\
    &\left. \times \left[ 
      1 - \cfrac{\mu^2 \bigl(2 R_{gg} + c_B (q_3 - 1) \bigr) 
        \left(e^{(2 R_{gg} + c_B (q_3 - 1))\frac{z_h^2}{2}}\Tilde{I}_1(z_h) -
          \Tilde{I}_2(z_h) \right)}{L^2 \left(1 
          - e^{(2R_{gg}+c_B(q_3-1))\frac{z_h^2}{2}}
        \right)^2} \right] \right|, 
    \\
    s = \ & \cfrac{1}{4} \left( \cfrac{L}{z_h} \right)^{1+\frac{2}{\nu}}
    e^{-(2R_{gg}-c_B)\frac{z_h^2}{2}-3 (p-c_B \, q_3) z_h^4}.
  \end{split} \label{eq:4.48}
\end{gather}

\begin{figure}[b!]
  \centering
  \includegraphics[scale=1]{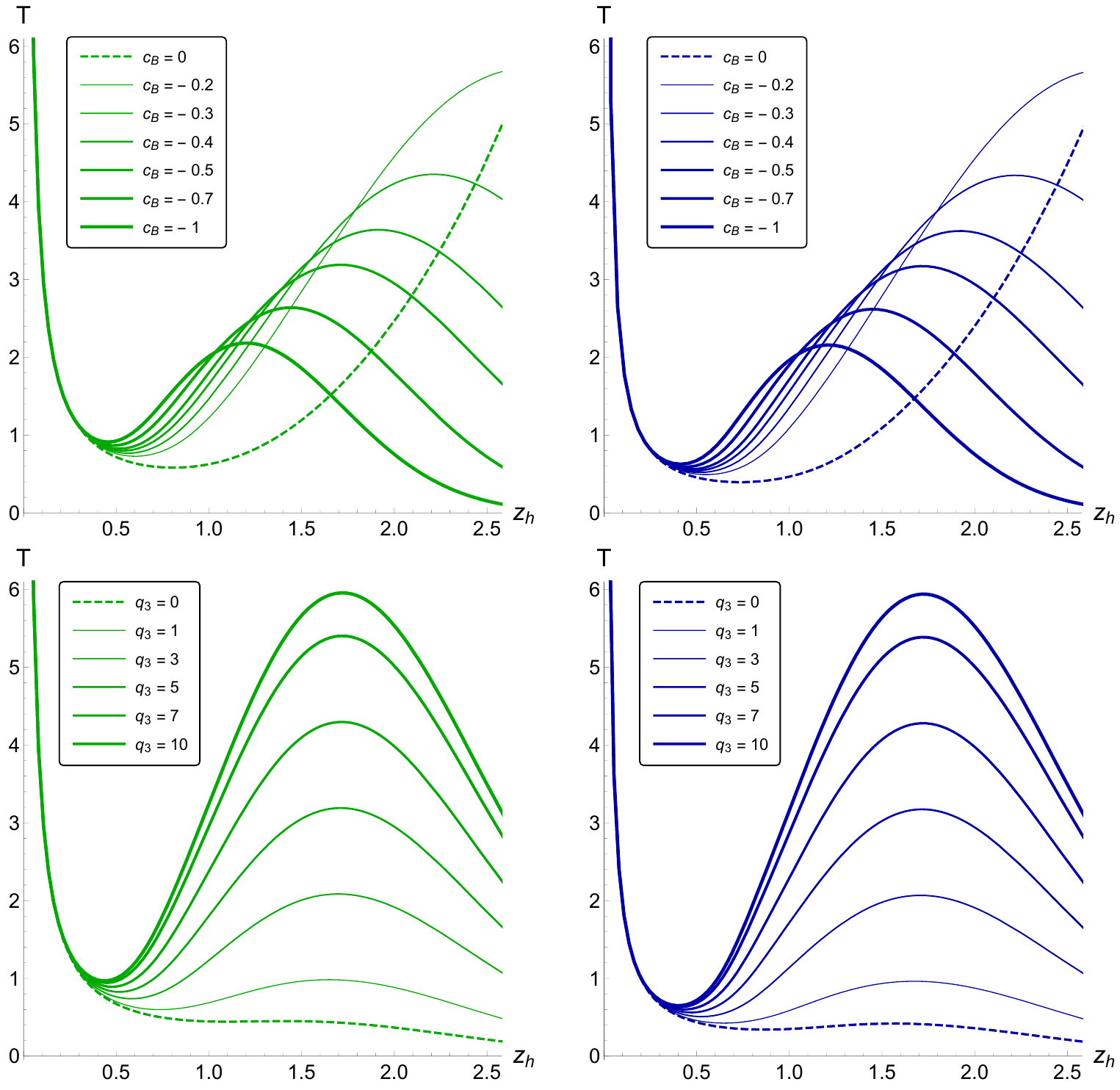} \\
  A \hspace{210pt} B
  \caption{Temperature $T(z_h)$ for different $c_B$, $q_3 = 5$ (1-st
    line) and for different $q_3$, $c_B = -\,0.5$  (2-nd line); $\nu =
    1$ (A) and $\nu = 4.5$ (B); $R_{gg} = 1.16$, $p = 0.273$, $\mu =
    0$; $[T] = [z_h]^{-1} =$~GeV.}
  \label{Fig:Tvszhmu0}
\end{figure}

\begin{figure}[t!]
  \centering
  \includegraphics[scale=1]{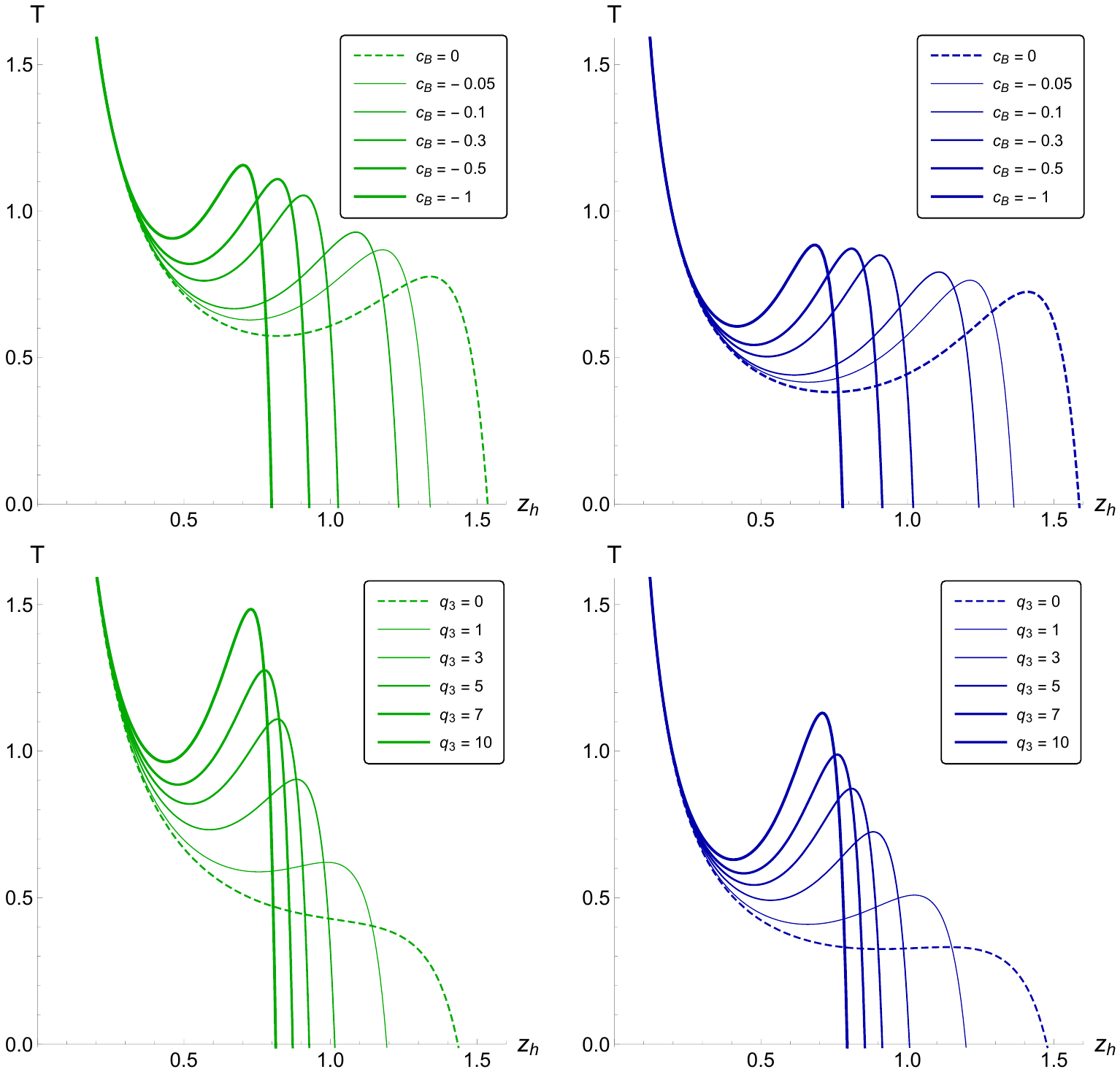} \\
  A \hspace{210pt} B
  \caption{Temperature $T(z_h)$ for different $c_B$, $q_3 = 5$ (1-st
    line) and for different $q_3$, $c_B = -\,0.5$  (2-nd line); $\nu =
    1$ (A) and with $\nu = 4.5$ (B); $R_{gg} = 1.16$, $p = 0.273$,
    $\mu = 0.3$; $[T] = [z_h]^{-1} =$~GeV.}
  \label{Fig:Tvszhmu03}
\end{figure}

The behavior of the temperature $T$ as a function of the horizon
radius $z_h$ for different values of the magnetic coefficient $c_B$
(1-st line) and the magnetic ``charge'' $q_3$ (2-nd line) in
backgrounds with different primary anisotropy $\nu = 1$ (A) and $\nu =
4.5$ (B) and zero chemical potential $\mu = 0$ is shown in
Fig.\ref{Fig:Tvszhmu0}. The system temperature is obviously sensible
to the magnetic field parameters in such a way that the temperature
minimum grows with the magnetic coefficient $c_B$ absolute value
(Fig.\ref{Fig:Tvszhmu0}, 1-st line) and the magnetic ``charge'' $q_3$
(Fig.\ref{Fig:Tvszhmu0}, 2-nd line). Naively, it means that the
Hawking-Page phase transition temperature $T_{HP}$ value increases
with the magnetic field in all the cases considered. Note that to
investigate the  magnetic field effect on the phase transition
temperature below, we consider the magnetic coefficient $c_B$,
although  the magnetic ``charge'' $q_3$ acts on the transition
temperature in a similar way. As to the primary anisotropy, it makes
the temperature local minimum to decrease both for various $c_B$ and
$q_3$. Therefore, one can  expect the Hawking-Page phase transition
temperature $T_{HP}$ to increase during the isotropisation
process. But to check this point and the BB 
phase transition explicitly, we need to calculate the free energy of
the system.

\begin{figure}[t!]
  \centering
  \includegraphics[scale=1]{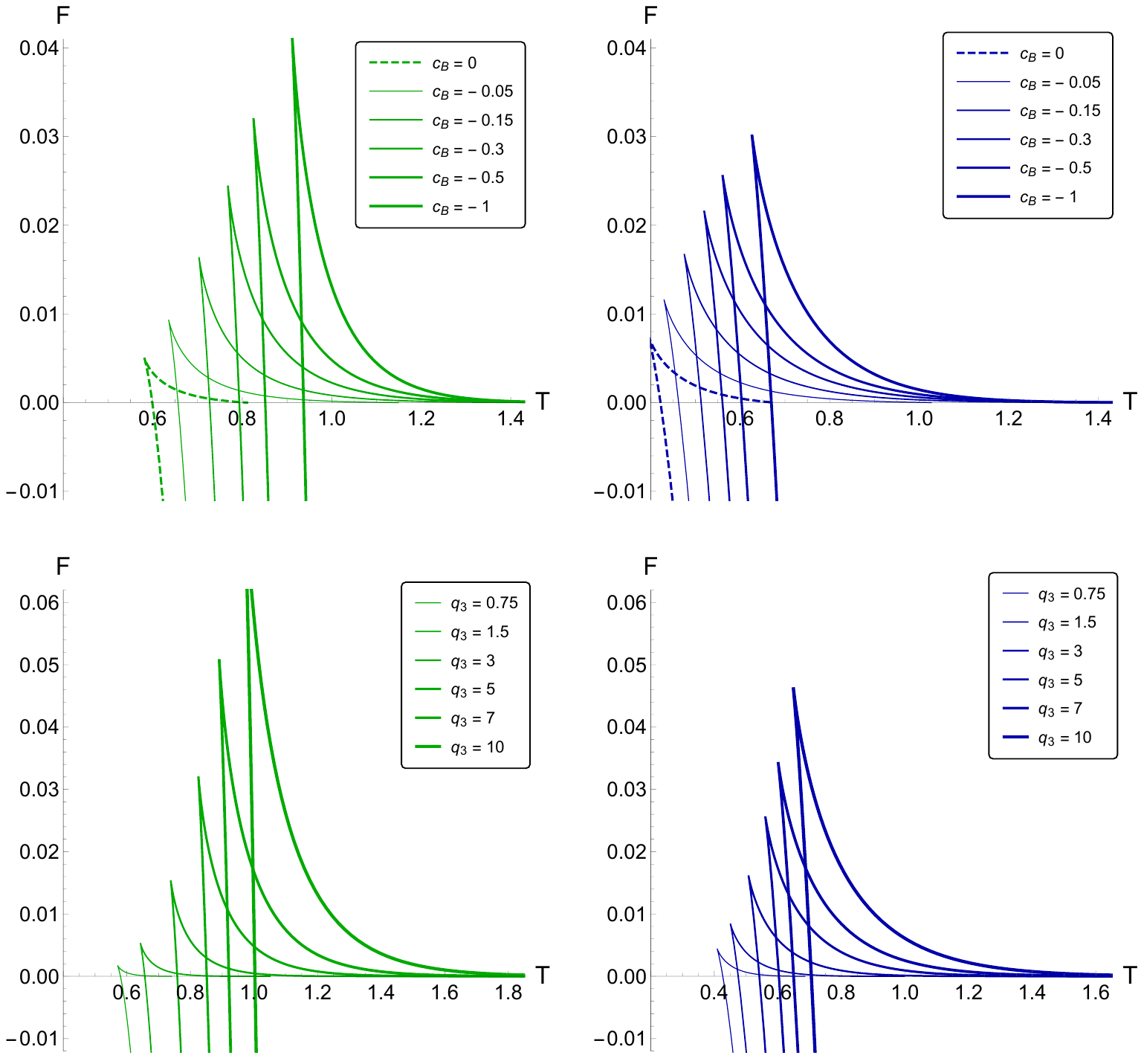} \\
  A \hspace{210pt} B
  \caption{Free energy 
  $F(T)$ for different $c_B$, $q_3 = 5$ (1-st line) and for different
  $q_3$, $c_B = - \, 0.5$ (2-nd line); $\nu = 1$ (A) and $\nu = 4.5$
  (B); $R_{gg} = 1.16$, $p = 0.273$, $\mu = 0$; $[F] = [T] =$ GeV.}
  \label{Fig:Fvstmu0}
\end{figure}

Now, we consider the non-zero chemical potential as we need to include
matter to investigate the realistic QGP with high baryonic density. In
Fig.\ref{Fig:Tvszhmu03} the system temperature $T$ in terms of the
horizon $z_h$ at the chemical potential $\mu = 0.3$ for different
values of $c_B$ (1-st line) and $q_3$ (2-nd line) is plotted. The
temperature is still very sensible to both magnetic field parameters,
and the Hawking-Page phase transition temperature $T_{HP}$
increases. Primary anisotropy makes the temperature local minimum to
decrease again and therefore the Hawking-Page phase transition
$T_{HP}$ requires less energy. To investigate the process precisely we
need to plot a free energy for different chemical potential values.

\begin{figure}[t!]
  \centering
  \includegraphics[scale=1]{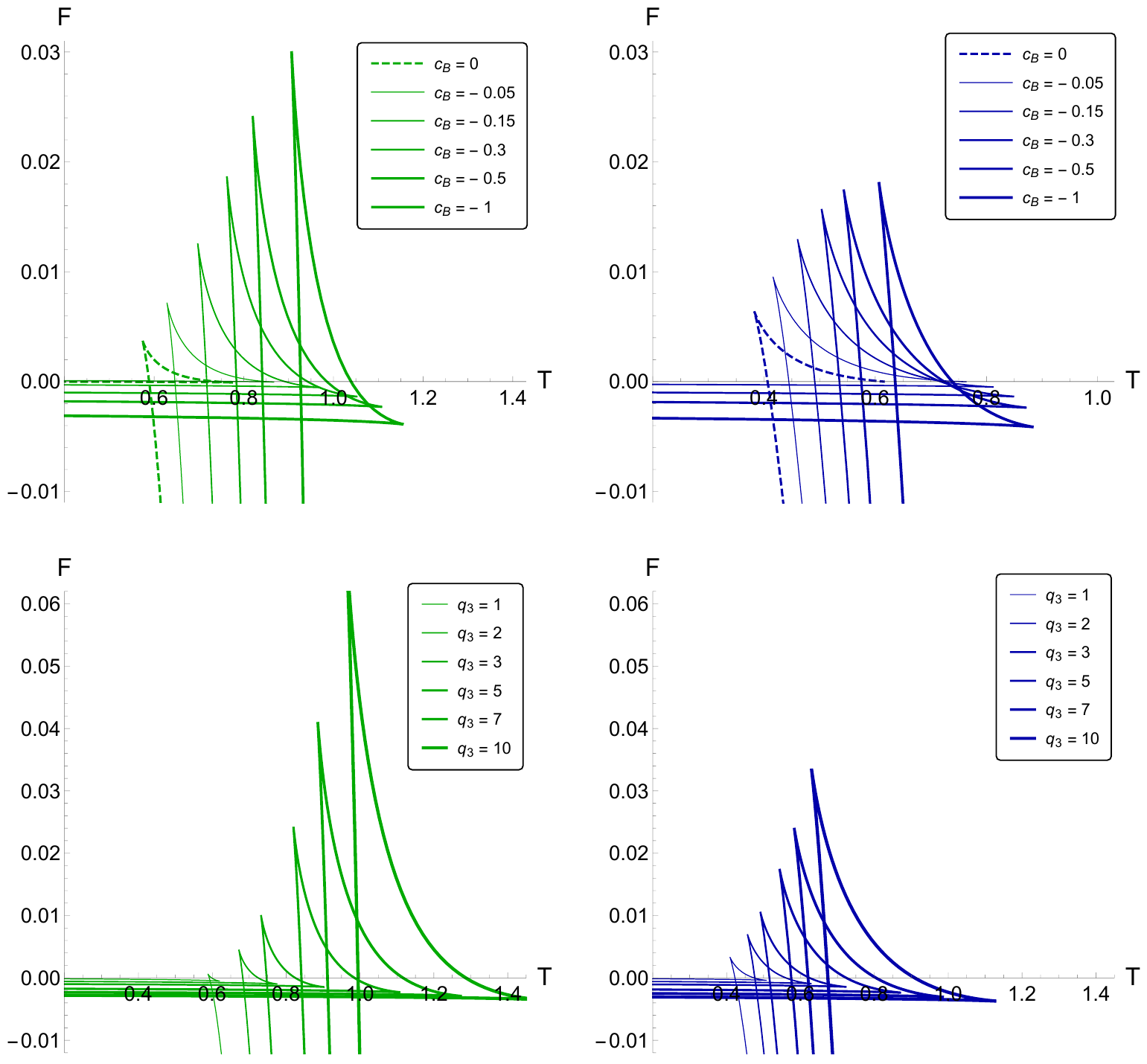} \\
  A \hspace{210pt} B
  \caption{Free energy $F(T)$ for different $c_B$, $q_3 = 5$ (1-st
    line) and $q_3$, $c_B = - \, 0.5$  (2-nd line); $\nu = 1$ (A) and
    $\nu = 4.5$ (B); $R_{gg} = 1.16$, $p = 0.273$, $\mu = 0.3$, in
    units $[F] = [T] =$ GeV.}
  \label{Fig:Fvstmu03}
\end{figure}

\subsubsection{Free energy and magnetic catalysis}

To investigate the first order phase transition we start from the free
energy $F$ as a function of temperature $T$ consideration for
different values of the magnetic field parameters $c_B$ and $q_3$ at
zero chemical potential. Regardless of the primary background
anisotropy ($\nu = 1$ in Fig.\ref{Fig:Fvstmu0}.A and $\nu = 4.5$ in
Fig.\ref{Fig:Fvstmu0}.B) the free energy is a multivalued function of
temperature and at $\mu = 0$ has two branches. One of them is positive
but asymptotically decreasing to zero, the other goes down into the
region of negative values sharply, almost vertically. The negative
part of this branch describes the large stable black hole, positive
free energy values correspond to the small unstable black hole and at
$F = 0$ thermal gas is formed. Temperature of the phase transition
from a small black hole to thermal gas known as Hawking-Page phase
transition is so called Hawking-Page temperature, $F(T_{HP}) =
0$. According to the holographic dictionary this process corresponds
to the first order phase transition in the dual 4-dim gauge theory. To
reveal the magnetic field effect on the free energy behavior and,
consequently, on the phase transition temperature $T_{HP}$, we
increase the absolute values of the magnetic parameters $c_B$
(Fig.\ref{Fig:Fvstmu0}, 1-st line) and $q_3$ (Fig.\ref{Fig:Fvstmu0},
2-nd line). In both cases $T_{HP}$ grows, thus the magnetic catalysis
phenomenon takes place for any primary anisotropy, but in the
background with higher primary anisotropy phase transition requires
lower temperature, like it was in previous works \cite{ARS-Heavy-2020,
  ARS-Light-2022, Arefeva:2022bhx, Arefeva:2021mag, Rannu:2022fxw,
  Arefeva:2023ter}.

\begin{figure}[b!]
  \centering
  \includegraphics[scale=1]{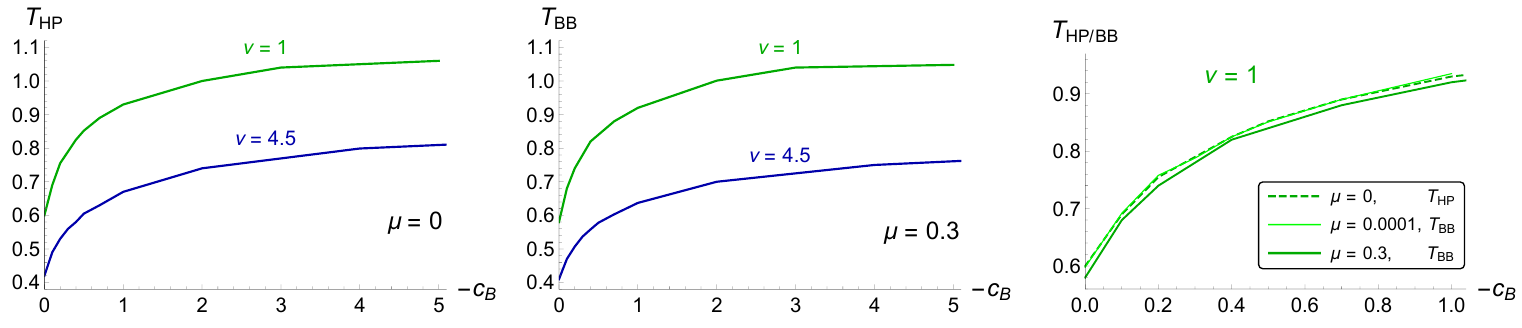} \\
  \quad A \hspace{130pt} B \hspace{130pt} C
  \caption{$T_{HP}$ as a function of $c_B$, $\mu = 0$ (A), $T_{BB}$ as
    a function of $c_B$, $\mu = 0.3$ (B) and their comparison (C);
    $\nu = 1$ (green lines) and $\nu = 4.5$ (blue lines); $R_{gg} =
    1.16$, $p = 0.273$, $q_3 = 5$; $[T] = [c_B]^{1/2} =$ GeV.}
  \label{Fig:TvscBmc}
\end{figure}

At a non-zero chemical potential there appears the free energy third
(almost horizontal) branch, and a closed structure called a
``swallow-tail'' appears (Fig.\ref{Fig:Fvstmu03}, $\mu = 0.3$). Its
self-intersection point describes  the first order phase transition,
and its temperature is marked as BB temperature $T_{BB}$. All other
tendencies preserve: magnetic field parameters $c_B$ and $q_3$ cause
the phase transition growth while the primary anisotropy $\nu$ lowers
it. This allows us to expect the MC-effect to be general on the first
order phase diagram. All these results are summarized in
Fig.\ref{Fig:TvscBmc}, where the phase transition temperature
dependence on the magnetic coefficient $c_B$ for $\mu = 0$ (A) and
$\mu = 0.3$ (B) in the background with different primary anisotropy is
depicted; the results for various chemical potentials in primary
isotropic case are confronted in Fig.\ref{Fig:TvscBmc}.C.

\begin{figure}[b!]
  \centering
  \includegraphics[scale=1]{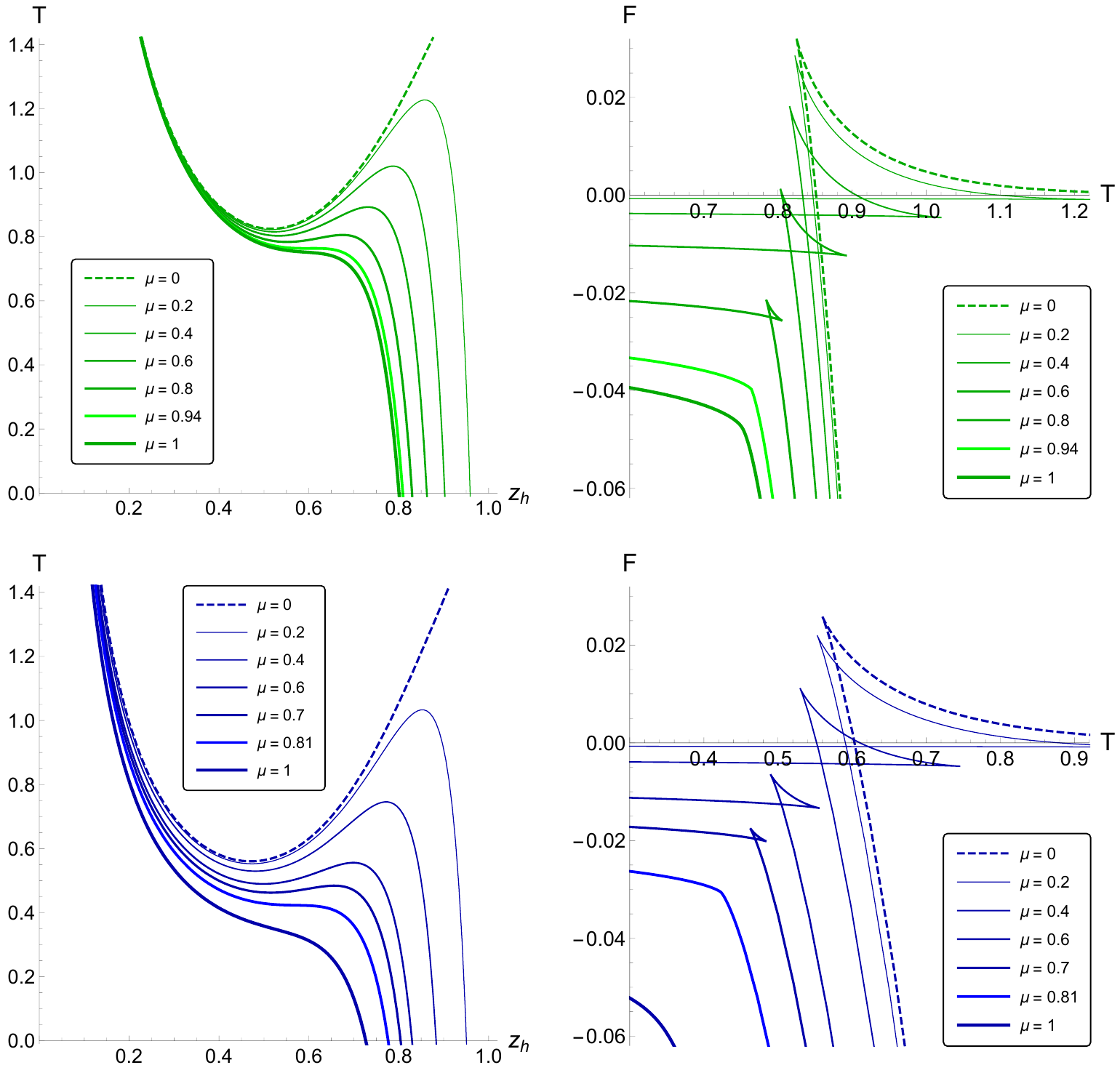} \\
  A \hspace{220pt} B
  \caption{Temperature $T(z_h)$ for different $\mu$ (A) and  free
    energy $F(T)$ for different $\mu$ (B); $\nu = 1$ (1-st line) and
    $\nu = 4.5$ (2-nd line); $R_{gg} = 1.16$, $p = 0.273$, $q_3 = 5$
    and $c_B = -\,0.5$, in units $[T] = [z_h]^{-1} =$ GeV.}
  \label{Fig:TFvstmudiff}
\end{figure}

To obtain the full picture of the first order phase transition the
free energy and consequently the temperature dependence on the
chemical potential should be considered. At $\mu = 0$ the temperature
has minimum $T_{min} = T(z_{min})$
(Fig.\ref{Fig:TFvstmudiff}.A). Black hole solutions with $z_h >
z_{min}$  (small black holes) are unstable, and the essence of the
phase transition lies in the collapse of such a black hole into a
stable black hole with the same temperature, but $z_h < z_{min}$
(large black holes). There is no black hole solution for $T <
T_{min}$, therefore Hawking-Page phase transition occurs at $T_{HP}
\ge T_{min}$. At the very point $T_{min}$ a large stable black hole
transforms into the thermal gas. This point is clearly seen on the
free energy plot as $F(T_{min}) = 0$ (Fig.\ref{Fig:TFvstmudiff}.B).

For $\mu > 0$ the function $T(z_h)$ also has a local maximum $T_{max}
= T(z_{max})$, where $z_{max} > z_{min}$
(Fig.\ref{Fig:TFvstmudiff}.A), and this leads to the ``swallow-tail''
appearance in the free energy plot again
(Fig.\ref{Fig:TFvstmudiff}.B). Temperature of the phase transition,
i.e. the collapse from a small unstable black hole to a large stable
black hole, can be determined as the temperature of the
self-intersection point in the ``swallow-tail'' base of the
$F(T)$-curve. As the chemical potential increases both $T(z_{min})$
and $T(z_{max})$ decrease so that the difference between them reduces
and eventually disappears at some critical value of the chemical
potential. For example, in the magnetic field with $c_B = -\,0.5$ for
$\nu = 1$ and $\nu = 4.5$ it happens at $\mu_{CEP_{HQ}} \approx 0.94$
and $\mu_{CEP_{HQ}} \approx 0.81$ correspondingly
(Fig.\ref{Fig:TFvstmudiff}.A). This process is reflected by the
``swallow-tail'' decrease on the free energy plot
(Fig.\ref{Fig:TFvstmudiff}.B).  For $\mu > \mu_{CEP_{HQ}}$ the black
hole temperature becomes a monotonic function of the horizon and its
free energy becomes smooth. Note that additional details on this
subject can be found in papers on previous considerations
\cite{AR-2018,ARS-Light-2020,ARS-Heavy-2020}.

\subsubsection{Phase diagrams}

\begin{figure}[b!]
  \centering
  \includegraphics[scale=1]{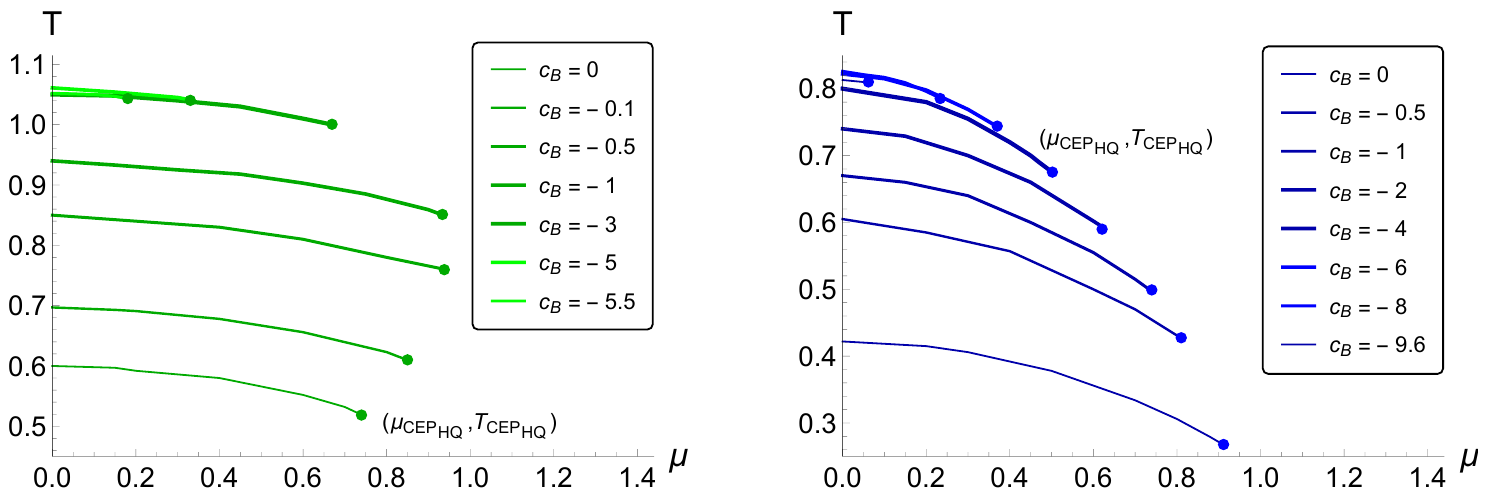} \\
  A \hspace{210pt} B \\
 \includegraphics[scale=1]{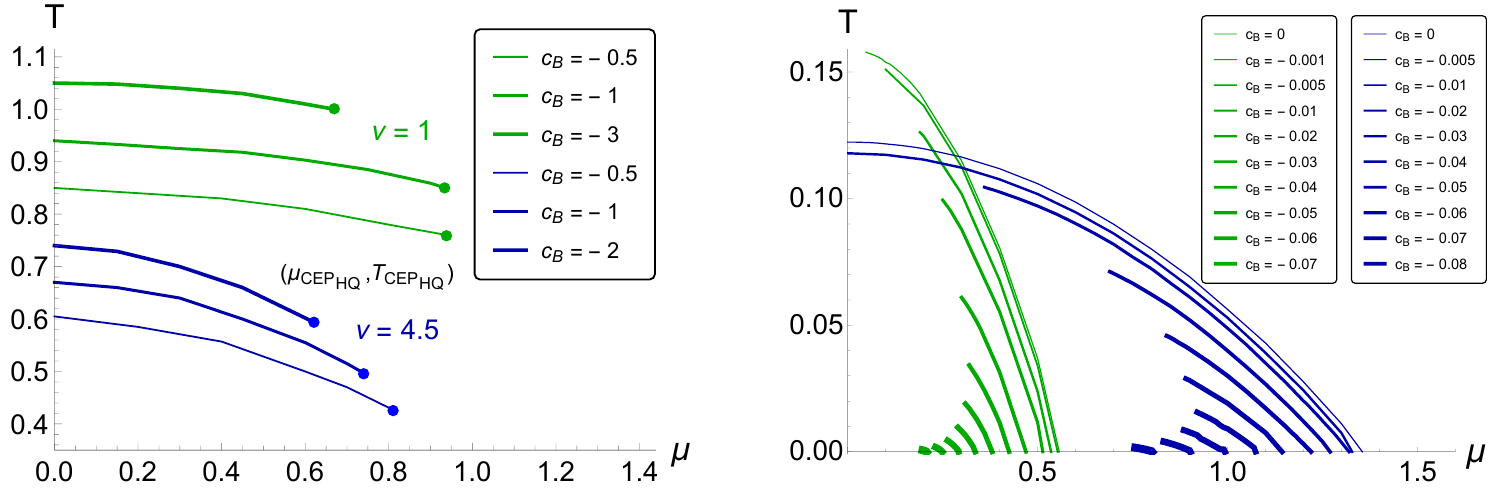} \\
  C \hspace{210pt} D
  \caption{The phase diagrams in the $(\mu,T)$-plane for heavy quarks
    with different $c_B$ for $\nu = 1$ (A) and $\nu = 4.5$ (B), the
    comparison between $\nu=1$ and $\nu=4.5$ (C); $R_{gg} = 1.16$, $p
    = 0.273$, $q_3 = 5$, in units $[T] = [\mu] =$ GeV; the phase
    diagrams in the $(\mu,T)$-plane for light quarks (D) from
    \cite{ARS-Light-2022}. In all plots we considered $\nu = 1$ (green
    lines) and $\nu = 4.5$ (blue lines) for different  $c_B$.}
  \label{Fig:Tvsmuheavyq}
\end{figure}

Phase diagram, i.e. the diagram of the confinement/deconfinement phase
transition displays the dependence of the phase transition temperature
 on the chemical potential. This phase diagram consists of two
 different phase transition lines, i.e. first order phase transition
 line and phase transition line for temporal Wilson loops  on the
 $(\mu$,$T)$-plane. In this research we considered  the first order
 phase transition only. In the ``heavy quarks'' version of our
 holographic model the phase transition line stretches over the
 interval $\mu \in [0, \mu_{max}]$, and the temperature in this
 interval drops (Fig.\ref{Fig:Tvsmuheavyq}). The rightmost point of
 the curve with coordinates $\bigl(\mu_{max}, T(\mu_{max})\bigr)$ is
 called a critical end point ($CEP_{HQ}$ for heavy quarks) and marks
 the free (not attached to the axis) end of the phase transition line
 (dots on the right end of the $T(\mu)$-curves). 

First of all, we really see the MC effect in the background with any
amount of anisotropy. For $\nu = 1$ CEP chemical potential
$\mu_{CEP_{HQ}}$ grows by increasing $|c_B|$ for the region $0<
\,|c_B| < \,0.5$,  after that $\mu_{CEP_{HQ}}$ decreases by increasing
$|c_B|$ and gets zero at $|c_B|=6$  when the phase transition line
completely disappears (Fig.\ref{Fig:Tvsmuheavyq}.A). For $\nu = 4.5$
$\mu_{CEP_{HQ}}$ decreases by increasing $|c_B|$, and the phase
transition line shortens until complete disappearence at $|c_B| \sim
\,10$, (Fig.\ref{Fig:Tvsmuheavyq}.B). We presented critical end points
for different $c_B$ with $\nu=1$ and $\nu=4.5$ in Table \ref{table}.

\begin{table}[t!]
  \centering
  \begin{tabular}{|l|c|l|c|}
    \hline
    \,\ $\nu=1$ & $(\mu_{CEP_{HQ}}, T_{CEP_{HQ}})$ &\,\ $\nu = 4.5$ &
    $(\mu_{CEP_{HQ}}, T_{CEP_{HQ}})$ \\ \hline
    $c_B=0$      & (0.74, 0.52) & $c_B=0$      & (0.91, 0.27) \\ \hline
    $c_B=-\,0.1$ & (0.85, 0.61) & $c_B=-\,0.5$ & (0.81, 0.43) \\ \hline
    $c_B=-\,0.5$ & (0.94, 0.76) & $c_B=-\,1$   & (0.74, 0.50) \\ \hline
    $c_B=-\,1$   & (0.93, 0.85) & $c_B=-\,2$   & (0.62, 0.59) \\ \hline
    $c_B=-\,3$   & (0.67, 1.00) & $c_B=-\,4$   & (0.50, 0.70) \\ \hline
    $c_B=-\,5$   & (0.33, 1.04) & $c_B=-\,6$   & (0.36, 0.75) \\ \hline
    $c_B=-\,5.5$ & (0.18, 1.04) & $c_B=-\,8$   & (0.23, 0.79) \\ \hline
    ---          & ---          & $c_B=-\,9.6$ & (0.06, 0.81) \\ \hline
  \end{tabular} 
  \caption{The critical end points for different $c_B$ with $\nu =
    1$ and $\nu = 4.5$}\label{table}
\end{table}

In both cases the $T_{HP/BB}$ rise slows down with increasing magnetic
field, before the complete disappearance of phase transition lines it
stops and even turns back. We observed that for the small region the
IMC appears at near-limit $c_B$ values, namely $|c_B| > \,5$ for $\nu
= 1$  and $|c_B| > \,8$ for $\nu = 4.5$ (light green and blue curves,
respectively). But it is obvious that in the primary isotropic
background the phase transition degenerates at a magnetic  field
weaker than in the primary anisotropic one. Besides for $\nu = 4.5$
phase transition temperature is lower and drops faster as $\mu$
increases (Fig.\ref{Fig:Tvsmuheavyq}.C). We can compare heavy quarks
phase diagrams (Fig.\ref{Fig:Tvsmuheavyq}.A,B,C) that describe MC and
light quarks phase diagram (Fig.\ref{Fig:Tvsmuheavyq}.D) that presents
IMC phenomenon \cite{ARS-Light-2022}.

\section{Conclusion and Discussion}\label{results}

In this research we studied the influence of the magnetic field on the
first order phase transition temperature. For this purpose we used the
``bottom-up'' approach and chose 5-dim Einstein-dilaton-Maxwell
holographic model with three Maxwell fields. In previous paper
\cite{ARS-Heavy-2020} only the IMC phenomenon was obtained. In this
research we look for a warp factor, that serves for a deformation
of metric, providing the MC phenomenon for the heavy quark model
\cite{ARS-Heavy-2020}. As our holographic model is phenomenological,
there is no systematic way to construct it.  We have to use the trial
and error method to get results compatible with existing experiments
and others theoretical methods.\\

Let us summarize our main results.

\begin{itemize}

\item  A new 5-dim exact analytical solution for  anisotropic
  holographic model of QGP is presented. One of its important features
  is inclusion of two types of anisotropy, caused by the spatial
  anisotropy, called primary, and the external magnetic field. Their
  influences on the background physical properties such as background
  phase transition are investigated.

\item Choosing the warp factor that deforms the metric is a key point
  of the current research. The warp factor $\fb(z) = e^{2{\cA}(z)} =
   e^{- \, c z^2/2 \, - \,2 p z^4}$ leads to IMC in the sense of
   decreasing of the critical temperature  with increasing the
   magnetic field $B$.

\item The MC effect is achieved for the warp factor $\fb(z) =  e^{
    - \, c z^2/2 \, - \,2 (p-c_B \, q_3) z^4}$ for $c_B$ not to be
  large (see discussion of the Figure \ref{Fig:Tvsmuheavyq}). This
  takes place for  zero chemical potential, i.e. for Hawking-Page-like
  (HP) phase transition, and for non-zero chemical potential,
  i.e. black hole-black hole first order. In both cases we get MC in
  the dual 4-dim gauge theory. 

\item The effect of primary anisotropy on black hole-black hole
    (BB) and Hawking-Page-like (HP) phase transition is investigated.
  It is found that anisotropy decreases both $T_{HP}$ and $T_{BB}$ for
  all values of magnetic field. 
  
\item The phase diagram, i.e. the dependence of the phase transition
  temperature on chemical potential, is built for different magnetic
  field magnitudes and different primary anisotropy values within the
  model constructed.

\item Complete disappearance of the phase transition lines for the
  primary isotropic background, $\nu = 1$, occurs at weaker magnetic
  field ($|c_B| \sim \,6$) in comparison to the anisotropic one, $\nu
  = 4.5$ ($|c_B| \sim \,10$).
    
\item Even for the near-limit chemical potential values the NEC to be
  preserved, as the consideration has sense and can be performed
  within the physical interval between the boundary and the second
  horizon in this model.
  
\item It is expected that for the presented model different quantities
  such as baryon density, entanglement entropy, electrical
  conductivity should have jump in the vicinity of the first order
  phase transition. This jump should strongly depend on the model
  parameters (anisotropy, magnetic field, chemical potential etc.)
  similar to previously considered models \cite{Arefeva:2021jpa,
    ARS-Light-2022, Arefeva:2020uec}. 

\item In \cite{ARS-Heavy-2020, AR-2018, ARS-2019, Arefeva:2021btm} the
  temporal and spatial Wilson loops were considered on the background
  for heavy quarks with two types of anisotropy with the warp factor
  $\fb(z) = e^{- \, c z^2/2}$. It would be interesting to study phase
  transition on this background with the new corrected warp
  factor. Also one can investigate energy loss and jet quenching on
  this background similar to \cite{Arefeva:2020bjk}. This will allow
  to obtain full confinement/deconfinement phase transition structure,
  that is determined by the Wilson loop and first order phase
  transition interplay.
\end{itemize}

In this paper we do not make calculations of the chiral condensate
$\langle\bar\psi \psi\rangle$. We would like to emphasize that to
calculate $\langle\bar\psi \psi\rangle$ one has to consider a new
action, i.e. chiral action including a few new fields $\chi$, and
solve corresponding equations of motion. We do not perform  these
calculations in the present paper, this will be the subject of the
future investigations. Similar  calculations have been performed in
\cite{Li:2020hau,Yang:2020hun,Bohra:2020qom} in different models.  The
chiral condensate $\langle\bar\psi \psi\rangle$ has been calculated
for light quark models in \cite{Li:2020hau,Yang:2020hun} for zero
magnetic field. The chiral condensate $\langle\bar\psi \psi\rangle$
has been calculated in \cite{Bohra:2020qom} for non zero $B$ for the
heavy quark holographic model with IMC in the sense of decreasing of
the critical temperature with increasing $B$ (i.e. without extra term
$z^4$ in the warp factor). It has been found that $\langle\bar\psi
\psi\rangle$ increasing with increasing $B$ that could be called MC in
the sense of the condensate value.

\ \\

\section{Acknowledgements}

The work of I.A. and A.H. was performed at the Steklov
International Mathematical Center and supported by the Ministry of
Science and Higher Education of the Russian Federation (Agreement
No.075-15-2022-265). The work of K.R. was performed within the
scientific project No. FSSF-2023-0003. K.R. and P.S. thank the
``BASIS'' Science Foundation (grants No.~22-1-3-18-1 and
No.~21-1-5-127-1, No.~23-1-4-43-1).

\appendix

\section{Equations of motion}\label{eeom}

Varying Lagrangian (\ref{eq:2.01}) over metric (\ref{eq:2.04}) we get
5 Einstein equations of motions: 
\begin{gather}
  \begin{split}
    {\bf 00}: \quad &- \left(\cfrac{L}{z}\right)^{3+\frac{2}{\nu}}
    e^{c_B z^2/2} \, \fb^{5/2} g \left[
      \cfrac{g'}{2} \left(
        \cfrac{3 \fb'}{2 \fb} 
        - \cfrac{2 + \nu \, (1 - c_B z^2)}{\nu z} \right)  \right. \\
    &+ g \left(
      \cfrac{3 \fb''}{2 \fb} - \cfrac{3 \fb'^2}{4 \fb^2}
      - \cfrac{3 \fb'}{2 \fb} \ \cfrac{2 - \nu c_B z^2}{\nu z}
      + \cfrac{3 + 2 \nu + \nu^2}{\nu^2 z^2} 
      - \cfrac{c_B \big( 3 - \nu \, (1 + c_B z^2) \big)}{\nu}
    \right)  \\
    &+ \left. \left(\cfrac{z}{L}\right)^2 \cfrac{f_0 A_t'^2}{4 \fb}
      + \left(\cfrac{L}{z}\right)^{2-\frac{4}{\nu}} 
      \cfrac{e^{-c_B z^2} f_1 q_1^2}{4 \fb}
      + \left(\cfrac{z}{L}\right)^{\frac{2}{\nu}} 
      \cfrac{f_3 q_3^2}{4 \fb}
      + \cfrac{g}{4} \, \phi'^2
      + \left(\cfrac{L}{z}\right)^2 \cfrac{\fb V}{2}
    \right] = 0,
  \end{split}\label{eq:A.05} \\
  \begin{split}
    {\bf 11}: \quad &- \left(\cfrac{L}{z}\right)^{3+\frac{2}{\nu}}
    e^{c_B z^2/2} \, \fb^{5/2} \left[
      \cfrac{g''}{2} + g' \left(
        \cfrac{3 \fb'}{2 \fb} 
        - \cfrac{2 + \nu \, (1 - c_B z^2)}{\nu z} \right)  \right. \\
    &+ g \left(
      \cfrac{3 \fb''}{2 \fb} - \cfrac{3 \fb'^2}{4 \fb^2}
      - \cfrac{3 \fb'}{2 \fb} \ \cfrac{2 - \nu c_B z^2}{\nu z}
      + \cfrac{3 + 2 \nu + \nu^2}{\nu^2 z^2} 
      - \cfrac{c_B \big( 3 - \nu \, (1 + c_B z^2) \big)}{\nu}
    \right)  \\
    &- \left. \left(\cfrac{z}{L}\right)^2 \cfrac{f_0 A_t'^2}{4 \fb}
      + \left(\cfrac{L}{z}\right)^{2-\frac{4}{\nu}} 
      \cfrac{e^{-c_B z^2} f_1 q_1^2}{4 \fb}
      - \left(\cfrac{z}{L}\right)^{\frac{2}{\nu}} 
      \cfrac{f_3 q_3^2}{4 \fb}
      + \cfrac{g}{4} \, \phi'^2
      + \left(\cfrac{L}{z}\right)^2 \cfrac{\fb V}{2}
    \right] = 0,
  \end{split}\label{eq:A.06} \\
  \begin{split}
    {\bf 22}: \quad &- \left(\cfrac{L}{z}\right)^{1+\frac{4}{\nu}}
    e^{c_B z^2/2} \, \fb^{5/2} \left[
      \cfrac{g''}{2} + g' \left(
        \cfrac{3 \fb'}{2 \fb} 
        - \cfrac{1 + \nu \, (2 - c_B z^2)}{\nu z} \right)  \right. \\
    &+ g \left(
      \cfrac{3 \fb''}{2 \fb} - \cfrac{3 \fb'^2}{4 \fb^2}
      - \cfrac{3 \fb'}{2 \fb} \ \cfrac{1 + \nu \, (1 - c_B z^2)}{\nu z}
      + \cfrac{1 + 2 \nu + 3 \nu^2}{\nu^2 z^2} 
      - \cfrac{c_B ( 2 - \nu c_B z^2)}{\nu}
    \right)  \\
    &- \left. \left(\cfrac{z}{L}\right)^2 \cfrac{f_0 A_t'^2}{4 \fb}
      - \left(\cfrac{L}{z}\right)^{2-\frac{4}{\nu}} 
      \cfrac{e^{-c_B z^2} f_1 q_1^2}{4 \fb}
      - \left(\cfrac{z}{L}\right)^{\frac{2}{\nu}} 
      \cfrac{f_3 q_3^2}{4 \fb}
      + \cfrac{g}{4} \, \phi'^2
      + \left(\cfrac{L}{z}\right)^2 \cfrac{\fb V}{2}
    \right] = 0,
  \end{split}\label{eq:A.07} \\
  \begin{split}
    {\bf 33}: \quad &- \left(\cfrac{L}{z}\right)^{1+\frac{4}{\nu}}
    e^{3 c_B z^2/2} \, \fb^{5/2} \left[
      \cfrac{g''}{2} + g' \left(
        \cfrac{3 \fb'}{2 \fb} 
        - \cfrac{1 + 2 \nu}{\nu z} \right)  \right. \\
    &+ g \left(
      \cfrac{3 \fb''}{2 \fb} - \cfrac{3 \fb'^2}{4 \fb^2}
      - \cfrac{3 \fb'}{2 \fb} \ \cfrac{1 + \nu}{\nu z}
      + \cfrac{1 + 2 \nu + 3 \nu^2}{\nu^2 z^2}
    \right)  
  \end{split} \label{eq:A.08}
\end{gather}
\begin{gather*}
  - \left. \left(\cfrac{z}{L}\right)^2 \cfrac{f_0 A_t'^2}{4 \fb}
  - \left(\cfrac{L}{z}\right)^{2-\frac{4}{\nu}} 
  \cfrac{e^{-c_B z^2} f_1 q_1^2}{4 \fb}
  + \left(\cfrac{z}{L}\right)^{\frac{2}{\nu}} 
  \cfrac{f_3 q_3^2}{4 \fb}
  + \cfrac{g}{4} \, \phi'^2
  + \left(\cfrac{L}{z}\right)^2 \cfrac{\fb V}{2}
    \right] = 0,  
\end{gather*} 
\begin{gather}
  \begin{split}
    {\bf 44}: \quad &- \left(\cfrac{L}{z}\right)^{3+\frac{2}{\nu}}
    e^{c_B z^2/2} \, \cfrac{\fb^{5/2}}{g} \left[
      \cfrac{g'}{2} \left(
        \cfrac{3 \fb'}{2 \fb} 
        - \cfrac{2 + \nu \, (1 - c_B z^2)}{\nu z} \right)  \right. \\
    &+ g \left(
      \cfrac{3 \fb'^2}{2 \fb^2}
      - \cfrac{3 \fb'}{2 \fb} \ \cfrac{2 + \nu \, (2 - c_B z^2)}{\nu z}
      + \cfrac{1 + 4 \nu + \nu^2}{\nu^2 z^2} 
      - \cfrac{1 + 2 \nu}{\nu} \ c_B
    \right)  \\
    &+ \left. \left(\cfrac{z}{L}\right)^2 \cfrac{f_0 A_t'^2}{4 \fb}
      + \left(\cfrac{L}{z}\right)^{2-\frac{4}{\nu}} 
      \cfrac{e^{-c_B z^2} f_1 q_1^2}{4 \fb}
      + \left(\cfrac{z}{L}\right)^{\frac{2}{\nu}} 
      \cfrac{f_3 q_3^2}{4 \fb}
      - \cfrac{g}{4} \, \phi'^2
      + \left(\cfrac{L}{z}\right)^2 \cfrac{\fb V}{2}
    \right] = 0.
  \end{split}\label{eq:A.09}
\end{gather}

We rewrite these  5 Einstein equations (remove factors in front of
(\ref{eq:A.05}-\ref{eq:A.09})) as: 
\begin{gather}
  \begin{split}
     {\bf [00]}:\quad &
    \cfrac{g'}{2} \left(
      \cfrac{3 \fb'}{2 \fb} 
      - \cfrac{2 + \nu \, (1 - c_B z^2)}{\nu z} \right)   \\
    &+ g \left(
      \cfrac{3 \fb''}{2 \fb} - \cfrac{3 \fb'^2}{4 \fb^2}
      - \cfrac{3 \fb'}{2 \fb} \ \cfrac{2 - \nu c_B z^2}{\nu z}
      + \cfrac{3 + 2 \nu + \nu^2}{\nu^2 z^2} 
      - \cfrac{c_B \big( 3 - \nu \, (1 + c_B z^2) \big)}{\nu}
    \right)  \\
    &+ \left(\cfrac{z}{L}\right)^2 \cfrac{f_0 A_t'^2}{4 \fb}
    + \left(\cfrac{L}{z}\right)^{2-\frac{4}{\nu}} 
    \cfrac{e^{-c_B z^2} f_1 q_1^2}{4 \fb}
    + \left(\cfrac{z}{L}\right)^{\frac{2}{\nu}} 
    \cfrac{f_3 q_3^2}{4 \fb}
    + \cfrac{g}{4} \, \phi'^2
    + \left(\cfrac{L}{z}\right)^2 \cfrac{\fb V}{2}
    = 0,
  \end{split}\label{eq:2.10} \\
  \begin{split}
    {\bf [11]}: \quad &
    \cfrac{g''}{2} + g' \left(
      \cfrac{3 \fb'}{2 \fb} 
      - \cfrac{2 + \nu \, (1 - c_B z^2)}{\nu z} \right)  \\
    &+ g \left(
      \cfrac{3 \fb''}{2 \fb} - \cfrac{3 \fb'^2}{4 \fb^2}
      - \cfrac{3 \fb'}{2 \fb} \ \cfrac{2 - \nu c_B z^2}{\nu z}
      + \cfrac{3 + 2 \nu + \nu^2}{\nu^2 z^2} 
      - \cfrac{c_B \big( 3 - \nu \, (1 + c_B z^2) \big)}{\nu}
    \right)  \\
    &-  \left(\cfrac{z}{L}\right)^2 \cfrac{f_0 A_t'^2}{4 \fb}
    + \left(\cfrac{L}{z}\right)^{2-\frac{4}{\nu}} 
    \cfrac{e^{-c_B z^2} f_1 q_1^2}{4 \fb}
    - \left(\cfrac{z}{L}\right)^{\frac{2}{\nu}} 
    \cfrac{f_3 q_3^2}{4 \fb}
    + \cfrac{g}{4} \, \phi'^2
    + \left(\cfrac{L}{z}\right)^2 \cfrac{\fb V}{2}
    = 0,
  \end{split}\label{eq:2.11} \\
  \begin{split}
    {\bf [22]}: \quad &
    \cfrac{g''}{2} + g' \left(
      \cfrac{3 \fb'}{2 \fb} 
      - \cfrac{1 + \nu \, (2 - c_B z^2)}{\nu z} \right)   \\
    &+ g \left(
      \cfrac{3 \fb''}{2 \fb} - \cfrac{3 \fb'^2}{4 \fb^2}
      - \cfrac{3 \fb'}{2 \fb} \ \cfrac{1 + \nu \, (1 - c_B z^2)}{\nu z}
      + \cfrac{1 + 2 \nu + 3 \nu^2}{\nu^2 z^2} 
      - \cfrac{c_B ( 2 - \nu c_B z^2)}{\nu}
    \right)  
  \end{split} \label{eq:2.12}  
\end{gather}
\begin{gather*}
    -  \left(\cfrac{z}{L}\right)^2 \cfrac{f_0 A_t'^2}{4 \fb}
    - \left(\cfrac{L}{z}\right)^{2-\frac{4}{\nu}} 
    \cfrac{e^{-c_B z^2} f_1 q_1^2}{4 \fb}
    - \left(\cfrac{z}{L}\right)^{\frac{2}{\nu}} 
      \cfrac{f_3 q_3^2}{4 \fb}
      + \cfrac{g}{4} \, \phi'^2
      + \left(\cfrac{L}{z}\right)^2 \cfrac{\fb V}{2} = 0,
\end{gather*}
\begin{gather}
    \begin{split}
      {\bf [33]}: \quad &
      \cfrac{g''}{2} + g' \left(
        \cfrac{3 \fb'}{2 \fb} 
        - \cfrac{1 + 2 \nu}{\nu z} \right)   \\
      &+ g \left(
        \cfrac{3 \fb''}{2 \fb} - \cfrac{3 \fb'^2}{4 \fb^2}
        - \cfrac{3 \fb'}{2 \fb} \ \cfrac{1 + \nu}{\nu z}
        + \cfrac{1 + 2 \nu + 3 \nu^2}{\nu^2 z^2}
      \right)  \\
      &- \left(\cfrac{z}{L}\right)^2 \cfrac{f_0 A_t'^2}{4 \fb}
      - \left(\cfrac{L}{z}\right)^{2-\frac{4}{\nu}} 
      \cfrac{e^{-c_B z^2} f_1 q_1^2}{4 \fb}
      + \left(\cfrac{z}{L}\right)^{\frac{2}{\nu}} 
      \cfrac{f_3 q_3^2}{4 \fb}
      + \cfrac{g}{4} \, \phi'^2
      + \left(\cfrac{L}{z}\right)^2 \cfrac{\fb V}{2} = 0,
    \end{split}\label{eq:2.13}  \\
    \begin{split}
      {\bf [44]}: \quad & 
      \cfrac{g'}{2} \left(
        \cfrac{3 \fb'}{2 \fb} 
        - \cfrac{2 + \nu \, (1 - c_B z^2)}{\nu z} \right) \\
      &+ g \left(
        \cfrac{3 \fb'^2}{2 \fb^2}
        - \cfrac{3 \fb'}{2 \fb} \ \cfrac{2 + \nu \, (2 - c_B z^2)}{\nu
          z}
        + \cfrac{1 + 4 \nu + \nu^2}{\nu^2 z^2} 
        - \cfrac{1 + 2 \nu}{\nu} \ c_B
      \right)  \\
      &+  \left(\cfrac{z}{L}\right)^2 \cfrac{f_0 A_t'^2}{4 \fb}
      + \left(\cfrac{L}{z}\right)^{2-\frac{4}{\nu}} 
      \cfrac{e^{-c_B z^2} f_1 q_1^2}{4 \fb}
      + \left(\cfrac{z}{L}\right)^{\frac{2}{\nu}} 
      \cfrac{f_3 q_3^2}{4 \fb}
      - \cfrac{g}{4} \, \phi'^2
      + \left(\cfrac{L}{z}\right)^2 \cfrac{\fb V}{2}
      = 0,
    \end{split}\label{eq:2.14}
\end{gather}
where $'= \partial/\partial z$.

We can see, that equations (\ref{eq:2.10}--\ref{eq:2.14}) have rather
complicated form on the one hand and include repeating combinations of
terms on the other hand. For further operating let us combine these
Einstein equations into the linear combinations, thus excluding the
repeating terms and concentrating on the specific details. To do this
we use the following receipt:
\begin{gather}
  \begin{split}
    {\bf (I)} \ \, \qquad  &[11] - [00], \\
    {\bf (II)} \ \qquad &[00] - [44], \\
    {\bf (III)} \qquad &[11] - [22], \\
    {\bf(IV)} \, \qquad & [11] - [33], \\
    {\bf(V)} \ \, \qquad &[22] + [44].
  \end{split}\label{eq:2.15}
\end{gather}
Together with the variations over the scalar field and first vector
field (Maxwell field that serves a non-zero chemical potential and for
which we have chosen the electric ansatz) we get the following EOMs:

\begin{gather}
  \begin{split}
    \phi'' + \phi' \left(
      \cfrac{g'}{g} + \cfrac{3 \fb'}{2 \fb} 
      - \cfrac{\nu + 2}{\nu z} + c_B z
    \right)
    &+ \left( \cfrac{z}{L} \right)^2 \cfrac{(A_t')^2}{2 \fb g} \ 
    \cfrac{\partial f_0}{\partial \phi}
    - \left( \cfrac{L}{z} \right)^{2-\frac{4}{\nu}} 
    \cfrac{e^{-c_Bz^2} \, q_1^2}{2 \fb g} \ 
    \cfrac{\partial f_1}{\partial \phi} \ - \\
    &- \left( \cfrac{z}{L} \right)^{\frac{2}{\nu}} 
    \cfrac{q_3^2}{2 \fb g} \ \cfrac{\partial f_3}{\partial \phi}
    - \left( \cfrac{L}{z} \right)^2 \cfrac{\fb}{g} \,
    \cfrac{\partial V}{\partial \phi} = 0,
  \end{split} \label{eq:2.16} \\
  A_t'' + A_t' \left(
    \cfrac{\fb'}{2 \fb} + \cfrac{f_0'}{f_0} 
    + \cfrac{\nu - 2}{\nu z} + c_B z
  \right) = 0, \label{eq:2.17} \\
  {\bf (I)} \qquad
  g'' + g' \left(
    \cfrac{3 \fb'}{2 \fb} - \cfrac{\nu + 2}{\nu z} + c_B z
  \right)
  - \left( \cfrac{z}{L} \right)^2 \cfrac{f_0 \,  (A_t')^2}{\fb}
  - \left(\cfrac{z}{L} \right)^{\frac{2}{\nu}} \cfrac{q_3^2 f_3}{\fb}
  = 0, \label{eq:2.18} \\
  {\bf (II)} \qquad  
  \fb'' - \cfrac{3 (\fb')^2}{2 \fb} + \cfrac{2 \fb'}{z}
  - \cfrac{4 \fb}{3 \nu z^2} \left(
    1 - \cfrac{1}{\nu}
    + \left( 1 - \cfrac{3 \nu}{2} \right) c_B z^2
    - \cfrac{\nu c_B^2 z^4}{2}
  \right)
  + \cfrac{\fb \, (\phi')^2}{3} = 0, \label{eq:2.19} \\
  \begin{split}
    {\bf (III)} \quad
    2 g' \left( 1 - \cfrac{1}{\nu} \right) 
    + 3 g \left( 1 - \cfrac{1}{\nu} \right) \left(
      \cfrac{\fb'}{\fb} - \cfrac{4 \left( \nu + 1 \right)}{3 \nu z}
      + \cfrac{2 c_B z}{3}
    \right)
    + \left( \cfrac{L}{z} \right)^{1-\frac{4}{\nu}} 
    \cfrac{L \, e^{-c_Bz^2} q_1^2 \, f_1}{\fb} = 0,
  \end{split} \label{eq:2.20} \\
  \begin{split}
    {\bf  (IV)} \quad
    2 g' &\left( 1 - \cfrac{1}{\nu} + c_B z^2 \right) 
    + 3 g \left[ \Big( 1 - \cfrac{1}{\nu} + c_B z^2 \Big) 
      \left(
        \cfrac{\fb'}{\fb} - \cfrac{4}{3 \nu z} + \cfrac{2 c_B z}{3}
      \right)
      - \cfrac{4 \left( \nu - 1 \right)}{3 \nu z} \right] + \\
    + &\left( \cfrac{L}{z} \right)^{1-\frac{4}{\nu}}
    \cfrac{L \, e^{-c_Bz^2} q_1^2 \, f_1}{\fb}
    - \left( \cfrac{z}{L} \right)^{1+\frac{2}{\nu}}
    \cfrac{L \, q_3^2 \, f_3}{\fb} = 0,
  \end{split}\label{eq:2.21} \\
  \begin{split}
    {\bf (V)} \quad 
    \cfrac{\fb''}{\fb} &+ \cfrac{(\fb')^2}{2 \fb^2}
    + \cfrac{3 \fb'}{\fb} \left(
      \cfrac{g'}{2 g} - \cfrac{\nu + 1}{\nu z} + \cfrac{2 c_B z}{3}
    \right)
    - \cfrac{g'}{3 z g} \left( 5 + \cfrac{4}{\nu} - 3 c_B z^2 \right)
    + \\
    &+ \cfrac{8}{3 z^2}
    \left( 1 + \cfrac{3}{2 \nu} + \cfrac{1}{2\nu^2} \right)
    - \cfrac{4 c_B}{3}
    \left( 1 + \cfrac{3}{2 \nu} - \cfrac{c_B z^2}{2} \right) 
    + \cfrac{g''}{3 g} 
    + \cfrac{2}{3} \left( \cfrac{L}{z} \right)^2 \cfrac{\fb V}{g} 
    = 0.  
  \end{split} \label{eq:2.22}
\end{gather}


\section{Coupling functions $f_1$, $f_3$ and dilaton potential
  $V$}\label{blackf}

We can obtain the exact form of the  coupling function $f_1$ that is
coupling function between the second Maxwell field $F_1$ and dilaton
field $\phi$ by utilizing the equation  (\ref{eq:2.29}) and inserting
the equation of $g(z)$ (\ref{eq:4.42}) and take into account its
derivative
\bea
 & g' (z) = - \frac{z^{\frac{2+\nu}{\nu}} \, e^{-\frac{c_B \, z^2}{2}
 +R_{gg}z^2+ 3(p-c_B  \, q_3 )z^4}}{\Tilde{I}_1(z_{h})}   \left[ 1- \frac{\mu^2 \, (c_B(-1+q_3)+2R_{gg})\left( \Tilde{I}_1(z_{h}) \, e^{\frac{1}{2} (c_B(-1+q_3)+2R_{gg}) z_h^2}- \Tilde{I}_2(z_{h})
 \right)}{(e^{\frac{1}{2} (c_B(-1+q_3)+2R_{gg}) z_{h}^2} -1)^2 \, L^2}
 \right] \nn   \\
& + \, 2\,  c_B \,  e^{c_Bz^2} \, z \left[1-\frac{\Tilde{I}_1(z)}{\Tilde{I}_1(z_h)}+ \frac{\mu^2 \, (c_B(-1+q_3)+2R_{gg})}{(e^{\frac{1}{2} (c_B(-1+q_3)+2R_{gg}) z_h^2} -1)^2 \, L^2} \left( \Tilde{I}_1(z_{h}) \Tilde{I}_2(z)- \Tilde{I}_1(z) \Tilde{I}_2(z_{h})  \right)    \right]  . 
   \label{eq:4.44}
\eea
After some algebra, one can obtain
\bea
&& f_1 (z) =-\frac{2(\nu-1)}{q_1^2 \, \nu^2 L^2} (\frac{L}{z})^{\frac{4}{\nu}} \,  e^{\frac{-2}{3}z^2 \, \left(-3 c_B + R_{gg}+ 3 (p-c_B \, q_3)z^2\right)} 
\Biggl[ -2 -2 \, \nu  \nn \\
&& + z^2 \,  \nu \left( 3c_B - 2 R_{gg} \, -12 (p-c_B \, q_3)z^2  + \frac{\mu^2 \, (c_B(-1+q_3)+2R_{gg}) \, z^\frac{2}{\nu} \, e^{\frac{1}{2} z^2 \, \left( 4R_{gg} +6 (p- c_B \, q_3)z^2 -4c_B \, +c_B \, q_3 \right)}}{(e^{\frac{1}{2} (c_B(-1+q_3)+2R_{gg}) z_h^2} -1)^2 \, L^2} \right) \nn \\
&& - \Biggl( \left( e^{\frac{1}{2} z^2 \, \left( -3c_B \, 2R_{gg} +6 (p- c_B \, q_3)z^2 \right)} \nu \, z^{2+\frac{2}{\nu}} + \left( -2 + \nu
\left( -2 + (3c_B - 2R_{gg})z^2 -12 \, (p-c_B \, q3)z^4  \right)\right) \Tilde{I}_1(z) \right) \nn \\
&& \times \left( \frac{1}{\Tilde{I}_1(z_h)} \,+ \frac{\mu^2 \, (c_B(-1+q_3)+2R_{gg})}{(e^{\frac{1}{2} (c_B(-1+q_3)+2R_{gg}) z_h^2} -1)^2 \, L^2} \, \frac{\Tilde{I}_2(z_h)}{\Tilde{I}_1(z_h)}
\right) \Biggr) \nn \\
&& +\, \frac{\mu^2 \, (c_B(-1+q_3)+2R_{gg}) \left( -2 + \nu
\left( -2 + (3c_B - 2R_{gg})z^2 -12 \, (p-c_B \, q_3)z^4  \right)\right) \Tilde{I}_2(z)}{(e^{\frac{1}{2} (c_B(-1+q_3)+2R_{gg}) z_h^2} -1)^2 \, L^2} 
 \Biggr], \label{eq:4.45}
\eea
where $\Tilde{I}_1(z)$  and $\Tilde{I}_2(z)$ were defined in  equations  \eqref{eq:4.43-1} and  \eqref{eq:4.43}.

The coupling function $f_3$ that is coupling function for the third Maxwell field $F_3$ dilaton field $\phi$ can be obtained by utilizing the equation  (\ref{eq:4.21}). It can be done by inserting the equation of $g'(z)$ (\ref{eq:4.44}) and take into account its derivative
\bea
 &&g'' (z) = e^{c_B \, z^2} (2c_B \,+ 4 c_B^2 \, z^2) \left[1- \frac{\Tilde{I}_1(z)}{\Tilde{I}_1(z_h)} + \frac{\mu^2 \, (c_B(-1+q_3)+2R_{gg})}{(e^{\frac{1}{2} (c_B(-1+q_3)+2R_{gg}) z_h^2} -1)^2 \, L^2} \left( \Tilde{I}_1(z_h) \Tilde{I}_2(z)- \Tilde{I}_1(z) \Tilde{I}_2(z_h)  \right)  \right] \nn \\ 
 &&-\,\frac{e^{-\frac{1}{2} z^2 \, (c_B-2 R_{gg}-6(p-c_B \, q_3)z^2)\,  z^{\frac{2}{\nu}} }}{\Tilde{I}_1(z_h)} 
 \Biggl[ \left(1- \frac{\mu^2 \, (c_B(-1+q_3)+2R_{gg})\left( \Tilde{I}_1(z_{h}) \, e^{\frac{1}{2} (c_B(-1+q_3)+2R_{gg}) z_h^2}- \Tilde{I}_2(z_{h})  \right)}{(e^{\frac{1}{2} c_B(-1+q_3)+2R_{gg}) z_{h}^2} -1)^2 \, L^2}  \right) \nn \\
&&\times \left( 1+\frac{2}{\nu} + (2 R_{gg}+3c_B)z^2 + 12 (p- c_B \, q_3)z^4\right) \nn \\&& -\, \frac{\mu^2 \, z^2 \, (c_B(-1+q_3)+2R_{gg}) \, e^{\frac{1}{2} (2R_{gg} + c_B(q_3-1)z^2)}}{(e^{\frac{1}{2} (c_B(-1+q_3)+2R_{gg}) z_h^2} -1)^2 \, L^2 } \Tilde{I}_1(z_h)
\Biggr].\label{eq:4.46}
\eea
For the coupling function $f_3$ we obtained

\bea
&& f_3(z) =- \frac{2 \, c_B \, e^{c_B\, z^2 -\frac{2R_{gg}}{3} \, z^2 -2(p- c_B \, q_3)z^4} (\frac{L}{z})^{\frac{2}{\nu}}}{\left(e^{\frac{1}{2} (c_B(-1+q_3)+2R_{gg}) z_h^2} -1 \right)^2 \, L^2 \, q_3^2 \, \nu \, \Tilde{I}_1(z_h)  } \times \Biggl[ \Bigl( e^{\frac{1}{2} z^2 \, \left( -3c_B \, 2R_{gg} +6 (p- c_B \, q_3)z^2 \, \right)} \, z^{2+\frac{2}{\nu}} \, \nu \nn \\&& + \, (-2+z^2 \, \nu \, (3c_B - 2R_{gg} - 12 (p-c_B \, q_3)z^2)) \, \Tilde{I}_1(z)  \Bigr) \Bigl( \left(e^{\frac{1}{2} (c_B(-1+q_3)+2R_{gg}) z_h^2} -1 \right)^2 \, L^2  \nn 
\\&& + \, \mu^2 (c_B(-1+q_3)+ 
  2R_{gg}) \, \Tilde{I}_2(z_h) \Bigr)   \Tilde{I}_1(z_h)  \Bigl( - e^{\frac{1}{2} z^2 \, \left( 4R_{gg} +6 (p- c_B \, q_3)z^2 -4c_B \, +c_B \, q_3 \right)} \, 
  \mu^2 (c_B(-1+q_3)+2R_{gg})\,\nn \\&& \times \, z^{2+\frac{2}{\nu}} \, \nu - 
\Bigl(2-z^2 \, \nu \, 
(3c_B - 2R_{gg} - 12 (p-c_B \, q_3)z^2) \Bigr) \nn \\&& \times \, \Bigl(
\left(e^{\frac{1}{2} (c_B(-1+q_3)+2R_{gg}) z_h^2} -1 \right)^2 \, L^2 + \mu^2 (c_B(-1+q_3)+2R_{gg})\, \Tilde{I}_2(z)
\Bigr)
 \Bigr)
\Biggr] \,.\label{eq:4.47}
 \eea  

The dilaton potential $V(z)$ can be obtained using equation (\ref{eq:4.81}) by plugging the blackening function $g(z)$
 (\ref{eq:4.43-1}) and the equations of $g'(z)$ (\ref{eq:4.44}) and $g''(z)$ (\ref{eq:4.46}). After some algebra we have
\bea
 && V(z) = \frac{e^{c_B \, z^2+\frac{2R_{gg}z^2}{3}+2(p-c_B\,q_3) z^4}}{2\,L^2\,\nu^2 \left(-1+e^{\frac{1}{2}(c_B(q_3-1)+2R_{gg})z_h^2} \right)^2}  \nn \\
 && \times \Biggl[ -e^{\frac{z^2}{2} (4R_{gg}+6p\,z^2+c_B(-4+q_3-6q_3\, z^2))} z^{2+\frac{2}{\nu}} \mu^2 \, \nu (c_B(q_3-1)+2R_{gg})(-2+\nu(-4+(c_B(3+q_3)-2R_{gg})z^2  \nn \\
&& -\,24(p-c_B\,q_3 )z^4)) -2 L^2 \left(-1+e^{\frac{1}{2}(c_B(q_3-1)+2R_{gg})z_h^2} \right)^2  \Bigl( 2+\nu \Bigl(6+4\nu+z^2 (-7c_B+6R_{gg}+36p\,z^2 \nn \\
&&-\,36c_B\,q_3\,z^2 +2\nu (-3c_B+2R_{gg}+(3c_B-2R_{gg})(c_B-R_{gg})z^2 -6(p-c_B\,q_3)(5c_B-4R_{gg})z^4\nn \\
&&+\,72(p-c_B\,q_3)z^6 ) ) \Bigr)
\Bigr) + \frac{2e^{\frac{-3}{2}c_B z^2(1+2q_3\, z^2) }}{\Tilde{I}_1(zh)} 
\Biggl(
-e^{R_{gg}\, z^2+3p\, z^4}\, z^{2+\frac{2}{\nu}}\, \nu\, \Biggl( 1+2\nu \Bigl(1+ \left(-c_B+R_{gg} \right) z^2 
\nn \\
&&+\,6  \left(p-c_B\,q_3 \right) z^4 \Bigr) 
+ \Tilde{I}_1(z) e^{\frac{3}{2}c_B z^2(1+2q_3\, z^2)}
\Biggl(2+\nu \Biggl(6+4\nu+z^2 \Bigl(-7c_B\,
+6R_{gg} +36p\, z^2 -36c_B \, q_3\, z^2 +2\nu
\nn \\
&&\times \left(-3c_B+2R_{gg}
+(3c_B-2R_{gg})(c_B-R_{gg})z^2 -6(p-c_B\,q_3)(5c_B-4R_{gg})z^4  +72(p-c_B\,q_3)z^6 \right) \Bigr)
\Biggr) \Biggr) \Biggr) 
\Biggr) \nn \\
&& \times
\Biggl( \Bigl(-1+e^{\frac{1}{2}(c_B(q_3-1)+2R_{gg})z_h^2} \Bigr)^2 \, L^2 +(c_B(-1+q_3)+2R_{gg})\, \mu^2 \, \Tilde{I}_2(zh)
\Biggr) -2  \mu^2 \, \Tilde{I}_2(z)(c_B(-1+q_3)
\nn \\
&&+\,2R_{gg}) \Biggl(2+\nu \Biggl(6+4\nu+z^2 \Biggl(-7c_B\, +6R_{gg} +36p\, z^2 -36c_B \, q_3\, z^2 +2\nu \Bigl(-3c_B+2R_{gg}+(3c_B-2R_{gg})
\nn \\
&&(c_B-R_{gg})z^2 -6(p-c_B\,q_3)(5c_B-4R_{gg})z^4 +72(p-c_B\,q_3)z^6 \Bigr) \Biggr) \Biggr) \Biggr)
 \Biggr].
   \label{eq:4.50}
\eea

\newpage

\section{Comparison with \cite{He:2020fdi}}\label{appC}

In this section we intend to compare the geometry of
\cite{ARS-Heavy-2020} with the metric and Lagrangian introduced in
\cite{He:2020fdi}:
\begin{gather}
  ds^2 = w_E^2 \left[
    - \, g(z) \, dt^2 + g_{11} dx_1^2 
    + g_{22} \left( dx_2^2 + dx_3^2 \right)
    + \cfrac{dz^2}{g(z)} \right] \! , \label{eq:4.04} \\
  g_{11} = e^{c_1 C(B) z^2}, \qquad 
  g_{22} = e^{c_2 C(B) z^2}, \label{eq:4.05} \\
  {\cL} = \sqrt{-g} \left[ R 
    - \cfrac{f(\phi)}{4} \, F^2
    - \cfrac{1}{2} \, \partial_{\mu} \phi \, \partial^{\mu} \phi
    - V(\phi) \right], \label{eq:4.06}  \\
  \phi = \phi(z), \nn \\
  \mbox{electro-magnetic ansatz $F$:} \quad
  A_{\mu} = \bigl( A_t(z),0,0,A_3(x_2),0 \bigr), \quad 
  \mu = 0,1,2,3,4. \label{eq:4.07}
\end{gather}
Components $F_{00} \sim A_t(z)$ and $F_{23} \sim A_3(x_2)$ of the
united electro-magnetic field $F_{\mu\nu}$ from \cite{He:2020fdi}
formally correspond to electric Maxwell field $F_{\mu\nu}^0$ and first
magnetic Maxwell field $F_{\mu\nu}^1$ from
\cite{ARS-Heavy-2020}. Magnetic component $F_{23}$ acts along the
$x_1$-direction in (\ref{eq:4.04}), but it is real magnetic field, not
an effective source of primary anisotropy $\nu$, as \cite{He:2020fdi}
describes an isotropic model with magnetic field. Magnetic field
(analogous to $F_{\mu\nu}$ from \cite{ARS-Heavy-2020}) has non-zero
component $F_{23}$, acts along the $x_1$-direction and influences
$g_{11}$ making it differ from $g_{22}$ and $g_{33}$. Therefore we can
say that for models \cite{He:2020fdi} and \cite{ARS-Heavy-2020}
indexes $1 \leftrightarrow 3$:
\begin{gather}
  e^{c_1 (B) z^2} \longleftrightarrow e^{c_B z^2}, \qquad
  c_2 = 0. \label{eq:4.08}
\end{gather}
We also see that
\begin{gather}
  w_E = \cfrac{L}{z} \, \sqrt{\fb(z)} \ \Rightarrow \ 
  w_E z^2 = \sqrt{\fb} \, z, \ L = 1. \label{eq:4.09}
\end{gather}
In \cite{He:2020fdi} the coupling function $f_0$ and the warp factor
are:
\begin{gather}
  f_0 = \cfrac{z^{-(R_{gg}+\frac{c_1 C}{2})z^2}}{w_E z^2}, \qquad
  w_E = \cfrac{e^{d(z)}}{z}, 
  \label{eq:4.10}
\end{gather}
where $d(z) = - \, (R_{gg}/3) \, z^2 - p z^4$, $R_{gg} = 1.16$, $p =
0.273$, $c_1 = - \, 1$. Therefore, via simple comparison with
\cite{ARS-Heavy-2020}  we have $c \longleftrightarrow 4 R_{gg}/3 = 4
\cdot 1.16/3 = 1.54(6)$. In addition for the coupling function $f_0$
we considered
\begin{gather}
  f_0 = e^{-(R_{gg}+\frac{c_B q_3}{2})z^2} \,
  \cfrac{z^{-2+\frac{2}{\nu}}}{\sqrt{\fb}} \label{eq:4.27}
\end{gather}
in this research.
We'll leave $c_B$ as a model parameter to preserve an opportunity to
fit magnetic field back reaction on the metric (\ref{eq:2.04}) and fix
AdS-radius to be $L = 1$ in all numerical calculations.

\end{document}